\documentclass[english,prl,twocolumn,nofootinbib]{revtex4-1}
\usepackage[T1]{fontenc}
\usepackage[latin9]{inputenc}
\usepackage{babel}
\usepackage{amstext}
\usepackage{graphicx}
\usepackage{esint}
\usepackage[unicode=true]
 {hyperref}
 \usepackage{caption}
 \usepackage{subcaption}

\makeatletter

\providecommand{\tabularnewline}{\\}

\makeatother

\begin{document}

\title{Principal Component Analysis of Diffuse \\Magnetic Neutron Scattering: a Theoretical
Study}

\author{Robert Twyman}

\author{Stuart J Gibson}

\author{James Molony}

\author{Jorge Quintanilla}
\email{j.quintanilla@kent.ac.uk}

\selectlanguage{english}%

\affiliation{School of Physical Sciences, University of Kent, Canterbury, Kent,
CT2 7NH, United Kingdom}

\thanks{JQ wishes to acknowledge useful discussions with S.T.~Carr, G.~M\"oller,
S.~Ramos, and T.~Tula.}
\begin{abstract}
We present a theoretical study of the potential of Principal Component Analysis
to analyse magnetic diffuse neutron scattering data on quantum materials.
To address this question, we simulate the scattering function $S\left(\mathbf{q}\right)$
for a model describing a cluster magnet with anisotropic spin-spin
interactions under different conditions of applied field and temperature.
We find high dimensionality reduction and that the algorithm can be
trained with surprisingly small numbers of simulated observations.
Subsequently, observations can be projected onto the reduced-dimensionality space defined by the learnt principal components. Constant-field
temperature scans correspond to trajectories in this space which
show characteristic bifurcations at the critical fields corresponding
to ground-state phase boundaries. Such plots allow the ground-state phase diagram to be accurately determined from finite-temperature
measurements.
\end{abstract}
\maketitle

\section{Introduction}

The study of quantum matter has emerged in recent decades as a major
field of scientific endeavour. The behaviour of many-body
systems is quite well understood at relatively high temperatures
where it is dominated by classical forces and entropy and where it
can be simulated efficiently using classical computers. However
quantum effects including entanglement and particle indistinguishability
make the equivalent, low-temperature problem much harder in principle.
Even so, a good understanding has emerged for various ordered ground
states, including the Landau Fermi liquid and states showing magnetic,
superconducting or topological order \citep{Bruus,Wen2013Mar}. Strongly-correlated
quantum matter \citep{Quintanilla2009PhysWorld}, on the other hand,
shows quantum correlations persisting at intermediate energy scales
and is less well understood with many outstanding questions. For example,
the precise relationship between the intermediate-temperature, ``liquid''
states and the various ground states in its phase diagrams remains unknown \citep{Laughlin2010Nov,Anderson2002}.

In recent years the arsenal available to tackle such challenging problems
has been enlarged by by the application of Machine Learning (ML).
For instance, artificial neural networks have been used to efficiently
encode the wave function of a many-body Hamiltonian, searching for
the ground state by reinforcement learning~\cite{ANN1}; to predict
the properties of one material from those of other substances, without
involving a model Hamiltonian~\cite{Conduit1}; and to detect phase transitions from piezoelectric relaxation measurements~\cite{Li2018Mar} and spectroscopic imaging scanning tunnelling microscopy~\cite{Zhang2019Jun}.

Here we propose an application of ML to magnetic neutron scattering
(NS). The neutron's intrinsic magnetic moment and availability of
neutron beams with wavelengths of the order of an angstrom make NS
one of the most powerful probes of magnetism in materials~\cite{Lovesey1987b}.
NS has provided, for instance: a thorough characterisation of the
magnetic excitation spectrum of the cuprates~\cite{Fong1999Apr,Dai1999May,Robert2006Nov,Vignolle2007Feb,Chan2016Mar};
strong evidence of magnetic monopole excitations in `spin ice' frustrated
magnets~\cite{Fennell2009Oct,Morris2009Oct}; and a quantitative understanding
of quantum phase transitions in magnetic insulators~\cite{Ruegg2003May,Lake2005Apr,Coldea2010Jan,Silverstein2018Apr}.

Our approach to magnetic NS is based on Principal Component Analysis
(PCA), a well-established technique for dimensionality reduction \citep{Pearson1901,Turk1991}
that can be regarded as a form of unsupervised machine learning. Formally,
PCA is equivalent to a linear autoencoder \citep{Geron2017} and is
a suitable initial step for a wide range of classification problems. In recent
years PCA and auto-encoders have been applied to data obtained through numerical simulation
of many-body systems \citep{Hu2017,Samarakoon2019}. It has been shown
that, when provided with detailed information on large representative
samples of microstates of such systems, such algorithms are capable of ``discovering''
important features in their phase diagrams, including order parameters
and phase transitions \citep{Hu2017}. On the other hand experimentally-accessible
information is normally limited and does not provide access to individual
microstates, consisting instead of thermal averages. The question
emerges: \emph{can PCA still identify important features from
such averages?}

Recently an autoencoder-based approach to magnetic diffuse neutron
scattering data on the ``spin-ice'' material Dy$_{2}$Ti$_{2}$O$_{7}$
has been demonstrated \citep{Samarakoon2019}. The autoencoder is
trained on a set of simulated neutron-scattering images. The simulations
correspond to a class of candidate model Hamiltonians. The trained
autoencoder is then used to describe real experimental data. It was
found that the autoencoder provides a compact description of the experimental
data, facilitating the identification of optimal model Hamiltonians,
and that it can also recognise distinct physical regimes in the simulations.
The latter suggests the question highlighted above can be answered
in the affirmative.

Here we ask whether PCA can be used to infer relevant features from
the data even in the absence of prior knowledge of a class of applicable
Hamiltonians (or in cases where the Hamiltonians might not be tractable).
This would require training the algorithm directly on the experimental
data. Given the scarcity of neutron flux, the training set would need
to consist of a limited number of scattering images, each with limited
resolution. The trained algorithm would then be used to filter additional
(but similarly limited) data sets and it would have to produce
qualitative signatures of any relevant features. Of particular interest
is the ability to infer ground-state phase boundaries from finite-temperature
data.

We perform a theoretical study to address the above questions, focusing on a class of models describing spin-1/2, anti-ferromagnetic,
ring-shaped molecular magnets. The field-dependent phase diagrams
of all instances of our model feature one or more level crossings
(LC) where the nature of the ground state changes discontinuously.
One of these LCs is a so-called ``entanglement transition'' (ET).
At the ET the ground state factorises exactly. These changes in the
system's ground state have clear signatures in simulated low-temperature,
high-resolution diffuse magnetic neutron scattering cross-sections
\cite{Irons2017}. The main question we tackle here is whether
a PCA can detect these features using a more limited number of lower-resolution
images. We will see that this is indeed possible. Moreover the ground-state
phase boundaries can be accurately determined from finite-temperature
data. In order to achieve all this we introduce the notion of a ``score bifurcation plot'' for principal components (PCs). This tool is only a small variation of the usual ``score plots'' in general use with PCA algorithms  but it shows in a particularly
transparent way how the different quantum ground states in the model
emerge from the high-temperature phase as the temperature is lowered.
We argue that this can be a useful tool for the identification of
quantum ground states from experimental data.

\section{\label{sec:Model}Model}
\begin{figure}
\includegraphics[width=0.90\columnwidth]{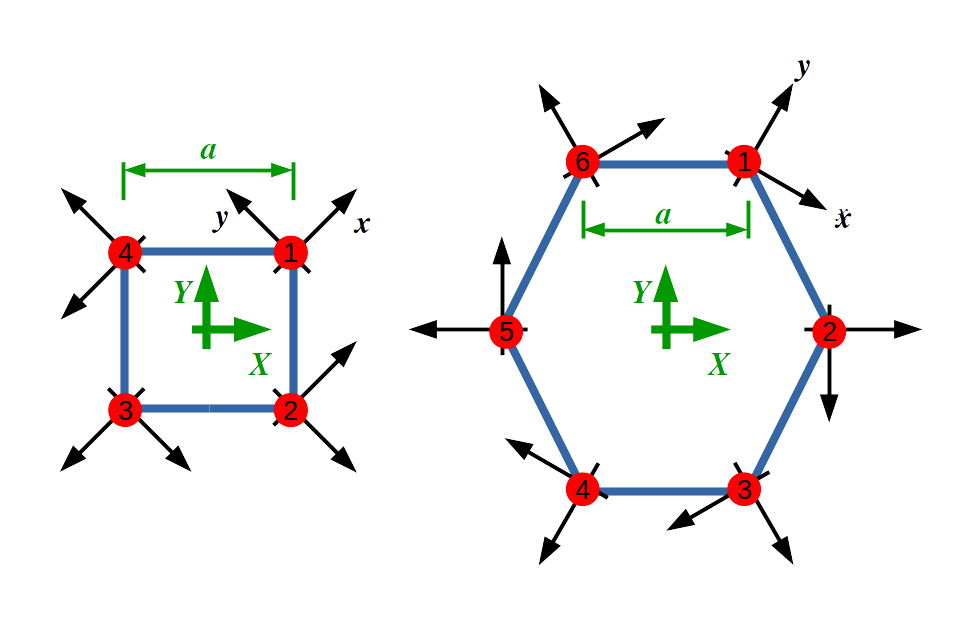}
\caption{\label{fig:model_pic}Our simple model of a planar, ring-shaped cluster magnet for the cases with $N=4$ (left) and $N=6$ (right) magnetic ions in the cluster. Reproduced from Ref.~\cite{Irons2017}.}
\end{figure}
\begin{widetext}
For the purpose of our study we consider a spin-1/2, anisotropic Heisenberg
ring in an applied magnetic field perpendicular to the plane of the
ring (for an illustration, see Fig.~\ref{fig:model_pic}). Assuming nearest-neighbour interactions only, the system has
the Hamiltonian
\begin{equation}
\hat{H}=\sum_{j=1}^{N}\left\{ -J\left[\left(1+\gamma\right)\hat{S}_{j}^{x}\hat{S}_{j+1}^{x}+\left(1-\gamma\right)\hat{S}_{j}^{y}\hat{S}_{j+1}^{y}+\Delta\hat{S}_{j}^{z}\hat{S}_{j+1}^{z}\right]-h\hat{S}_{j}^{z}\right\} .\label{eq:H}
\end{equation}
Here $N$ is the number of magnetic ions in the ring, which we assume
to be even, $J\text{ and }h$ are, respectively, the exchange and
field energies, and $\gamma$ and $\Delta$ are two dimensionless
parameters describing the anisotropy of the spin-spin interaction.
The operator $\hat{S}_{j}^{\alpha}$ represents the $\alpha^{\underline{\text{th}}}$
component of the spin at the $j^{\underline{\text{th}}}$ magnetic
site and the labels $x,y,z$ refer to the local mangetic axes at that
site. The $x$ and $y$ axes rotate from site to site so as to preserve
the $C_{N}$ rotational symmetry of the molecule around the $z$ axis,
which is fixed. Note that we have assumed that the interaction is
diagonal in this basis. An illustration of the geometry of the model
can be found in \citep[Ref.][Fig. 1]{Irons2017}. We take all three
components of the interaction to be anti-ferromagnetic, corresponding to $J<0$, and assume
without loss of generality $0\leq\gamma\leq1$ and $\Delta>0$ \footnote{Ignoring the spatial arrangement of the atoms, the Hamiltonian in
Eq.~(\ref{eq:H}) can correspond to a number of distinct universality
classes: Heisenberg for $\Delta=\gamma=0;$ XY for $\gamma=0\neq\Delta$;
and Ising for $\Delta=0\neq\gamma$.}. The boundary conditions are enforced by setting $N+1\equiv1$.
\end{widetext}

The behaviour of the model defined by Eq.~(\ref{eq:H}) has been
studied extensively \citep{giorgi_ground-state_2009,giorgi_erratum:_2009,k._barwinkel_ground-state_2000,barwinkel_quantum_2003,de_pasquale_x_2009,Irons2017}.
For fixed $J,\gamma,\Delta$ it has $N/2$ ground state degeneracies
at $h=h_{1},h_{2},\ldots,h_{N/2}=h_{f},$ where the last degeneracy
takes place at the $N-$independent factorisation field $h_{f}=J\sqrt{\left(1+\Delta\right)^{2}-\gamma^{2}}$.
The simulated diffuse magnetic neutron scattering function $S\left(\mathbf{q}\right)$
for the geometry under consideration and with the scattering vector
$\mathbf{q}$ within the plane of the molecule \citep{Irons2017}
shows qualitative changes from anti-ferromagnetic correlations for
$h<h_{f}$ to ferromagnetic ones for $h>h_{f}$. This is consistent
with an entanglement transition from anti-parallel Bell states to
parallel Bell states, respectively, known to take place at $h_{f}$
\citet{amico_divergence_2006}. Less striking, but well-defined changes
also occur at the other level crossings. Specifically, numerical evidence
for a jump of $S\left(0\right)$ in the ground state taking place
at all $N/2$ level crossings has been obtained for $N=4,6$ \citep{Irons2017},
8 and 10 \citep{Irons2016}. At finite temperatures the jumps become
crossovers which get smoother as the temperature is raised further.

The codes we used for this study are freely available as open source from Refs.~\citep{magneto2020} (neutron scattering simulations) and \citep{jarvis2020} (PCA). They require only the Octave computer language~\cite{OctaveCite}. Further details are given in the appendix.

\section{\label{sec:Dimensionality-reduction}Dimensionality reduction}

For any given set of values of the parameters of our model, the scattering
function $S\left(\mathbf{q}\right)$ mentioned above can be interpreted
as an image \footnote{Throughout this work we assume that any background terms have been
substracted from our scattering functions: $S\left(\mathbf{q}\right)\to S\left(\mathbf{q}\right)-\Omega^{-1}\int d\mathbf{q\,}S\left(\mathbf{q}\right),$where
$\Omega\equiv\int d\mathbf{q}.$}. Giving the parameters different values allows us to generate different
images which can be subject to PCA. Quite generally, the result of
a PCA of any set of images is a complete, orthogonal basis set that
can be used to reconstruct exactly, through linear superposition,
any of the images in the original training set. The advantage of this
new basis is that the PCs are ordered, with the first basis element
capturing the largest amount of variance within the original data
set, the second capturing the second largest amount, and so forth.
For images comprising solely random pixel values this would offer no advantage but if the images are strongly-correlated
then a very good approximation to all the images in the training set
can be obtained using just the first few PCs. PCA can thus be regarded
as a technique for dimensionality-reduction and this
forms the basis of its application to problems such as face recognition
\citep{Turk1991}. Its effectiveness relies on correlations within
the training set: if correlation is high (e.g. all images represent
human faces) then a small number $M$ of PCs can capture most of the
variance in the data set. The details of our PCA procedure are given in the appendix.
\begin{figure}
\begin{subfigure}[b]{0.85\columnwidth}
\centering
  \includegraphics[width=1\columnwidth]{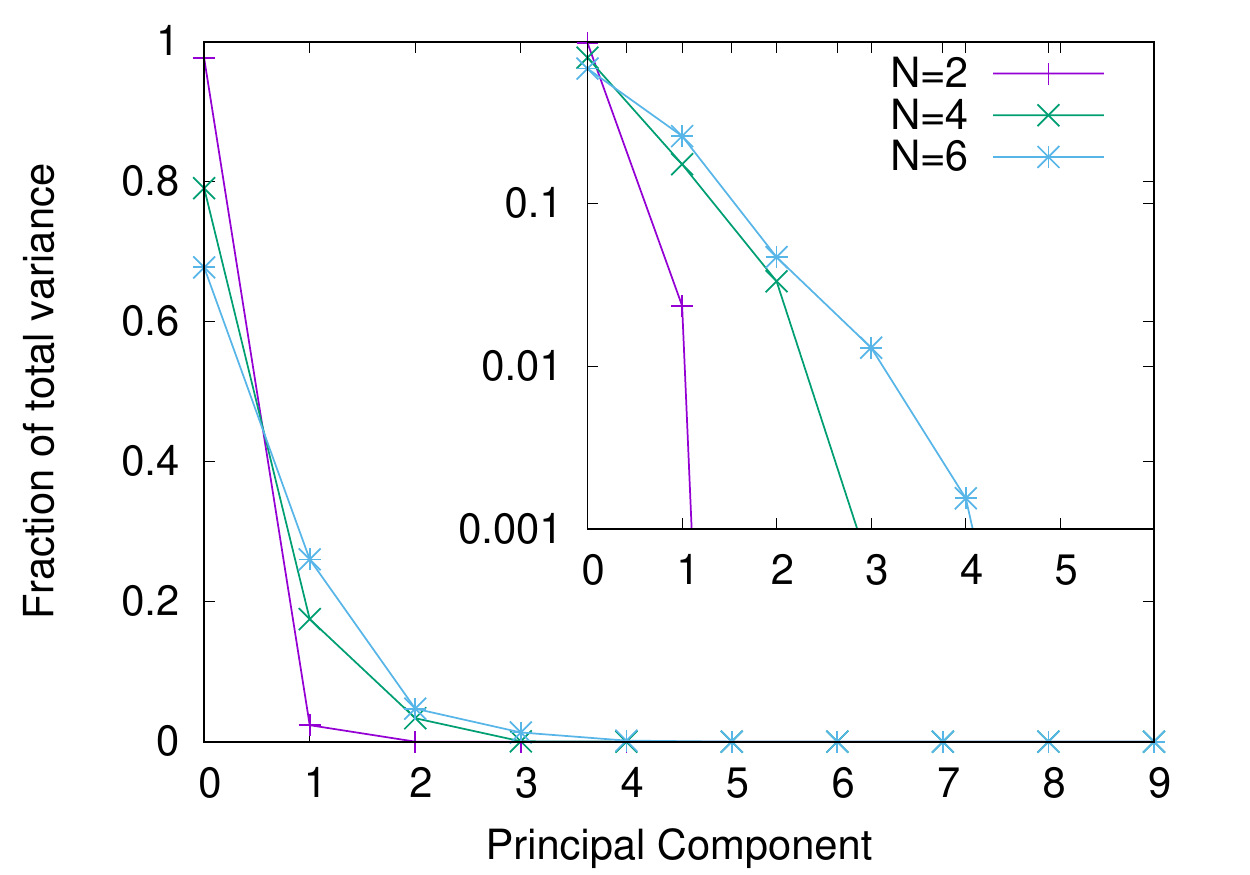}
  \caption{}
\end{subfigure}
\begin{subfigure}[b]{0.85\columnwidth}
\centering
  \includegraphics[width=1\columnwidth]{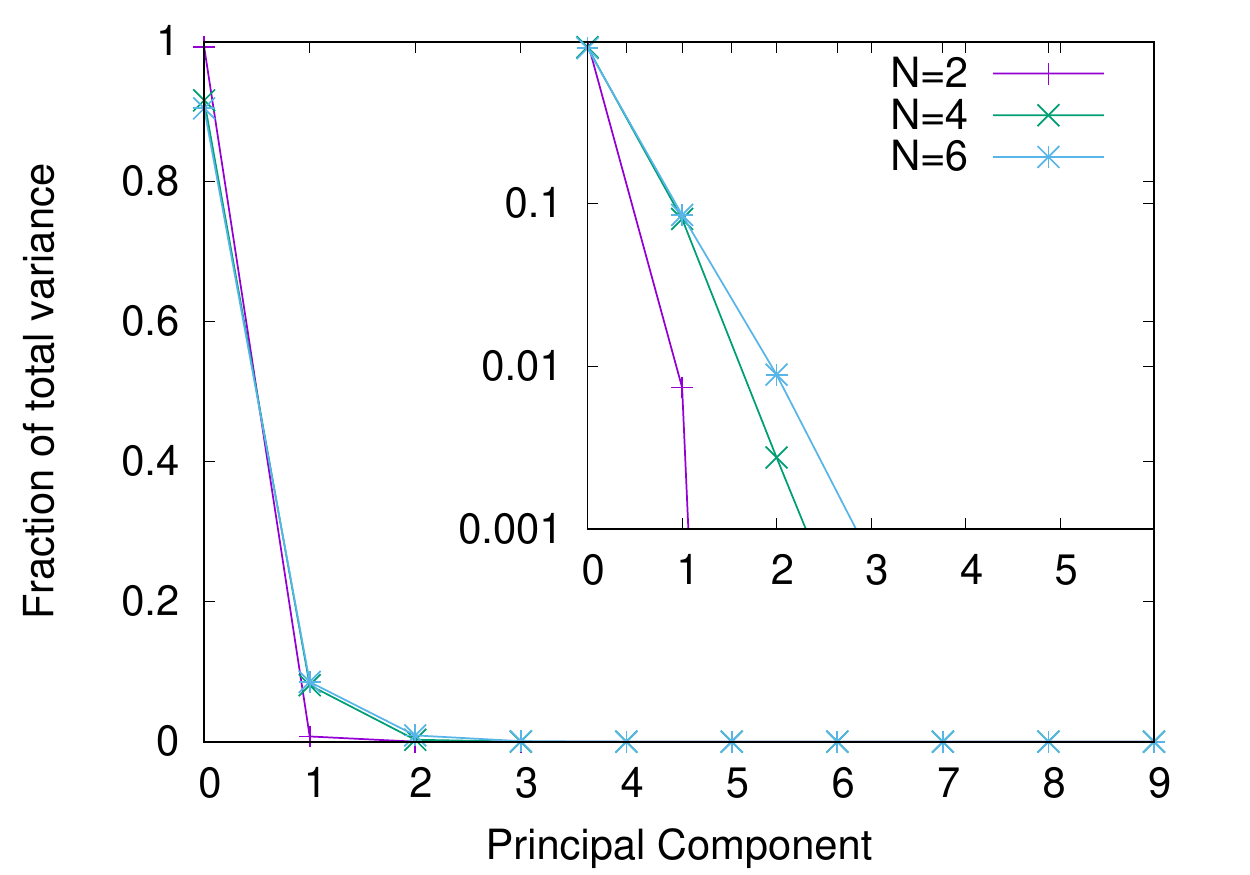}
  \caption{}
\end{subfigure}

\caption{\label{fig:scree_plots}Fraction of the variance in a set of simulated
diffuse magnetic neutron scattering images that is captured by the
first 10 PCs. Each curve was obtained by PCA of 500 random observations.
Each observation is a $24\times24$ pixel image obtained by computing
the scattering function $S\left(\mathbf{q}\right)$ of our cluster-magnet
model. A uniform mesh of $\mathbf{q}$-vectors with components $q_{x},q_{y}$
ranging from $-9\pi/a$ to $9\pi/a$, where $a$ is the distance between
nearest neighbours within the cluster, was used. Each curve corresponds
to a particular number $N$ of magnetic ions in the cluster, as indicated.
The insets show the same data on a logarithmic scale. (a) System parameters
$\gamma,\Delta,h,T$ varied randomly within the ranges $0<\gamma<1,\,0<\Delta<1,\,0<h<2J,$
and $0.01<T<2J$ (variation with respect to the exchange energy scale
$J$ is not necessary as it merely sets the overall energy scale).
(b) $h,T$ varied randomly within the same ranges as before with fixed
$\gamma=0.6\text{ and }\Delta=0$.}
\end{figure}

In our problem we expect to achieve significant reduction because
all images have been derived from instances of the same class of Hamiltonians.
Let us fix the number of magnetic moments in the molecule $N$ and
vary the parameters $\gamma,\Delta,h$ and temperature $T$ (all four
energies are measured in units of $J$). For each set of values we
can use the method in Refs.~\citep{Irons2016,Irons2017} to compute
$S\left(\mathbf{q}\right)$ for a fixed set of wave vectors $\mathbf{q}$.
This results in a set of images which can then be classified by a
standard PCA algorithm. Our expectation is borne out by the scree plots in Fig.~\ref{fig:scree_plots}~(a).
Specifically, we find that $M=4$ PCs suffice to capture 99\%
of the variance in the data set for the range of values of $N$ shown
in the graph.

In an experimental situation, we expect the parameters defining the
strength and anisotropy of magnetic interactions to be fixed for a
given material, while the strength of the externally-applied magnetic
field and temperature can vary. Scree plots for a representative case ($\gamma=0.6,\Delta=0$)
are shown in Fig.~\ref{fig:scree_plots}~(b).\footnote{\label{foot:gamma_choice_footnote}For our chosen value of the anisotropy parameters, $\gamma=3/5,\Delta=0$, we obtain the simple fraction $h_f=4/5$ exactly, which is convenient and motivates that choice; more generally $h$ is a real number but this choice does not introduce any qualitative differences.} We find that now $M=2$
captures 99\% of the variance in the training set.

Our results indicate that the number $M$ of PCs necessary to reproduce
to very high accuracy all the images in the training set is bound
by the number $M_{p}$ of free parameters used to generate the training
images. This might suggest an unbiased (model-independent) way to constrain
experimentally the number of independent parameters describing a class
of related materials -an important step in the derivation of a model
Hamiltonian. We note, however, that $M$ is more likely related to the number of distinct states in the phase diagram which, although related to $M_p$, can potentially be larger than it. Elucidating this will require the analysis of a broader range of models and is beyond the scope of the present work.

We highlight that unlike an autoencoder, where the number $M_{n}$
of neurons in the hidden layer has to be fixed \emph{a priori},\emph{
}our approach enables us to find out \emph{a posteriori }the number
$M$ of PCs needed to describe the data accurately. One would in principle
expect $M_{n}\sim M$ and thus for our model $M_{n}\sim M_{p}$. In
contrast, for the model of Ref.~\citet{Samarakoon2019} with $M_{p}=4$
dimensionless parameters $M_{n}=30$ was found to strike a good balance
between overfitting and underfitting.
This would suggest that in our case we achieve greater dimensionality
reduction.

Although PCA is equivalent to an autoencoder in the linear limit, we must note that the two studies are rather different. In \citet{Samarakoon2019} the aim is to automatically interpret structure factor data, from noisy measurements and to disambiguate among many possible solutions of the inverse scattering problem, whereas our approach is to achieve an embedding of the data that is amenable to human interpretation. Our work does not explicitly address experimental noise. However, we note that both autoencoder and PCA are known to have noise suppressing properties and therefore we anticipate that our method would also be applicable to experimental data.


\section{\label{sec:Size-of-training}Size of training set}

Once the value of $M$ has been set, other images not included in
the original training set can be accurately reconstructed using the
first $M$ PCs. For this to be possible, the following two conditions
must be met: i) the new images must be of the same type as those in
the training set; ii) the training set must be sufficiently representative.
The latter criterion will only be met if the number of images in
the training set $M_{t}$ is sufficiently large. This limits the feasibility
of obtaining training sets experimentally.



\begin{figure*}
    \centering
    \begin{subfigure}[b]{0.65\columnwidth}
        \centering
        \hspace{-0.85cm}\includegraphics[width=\columnwidth]{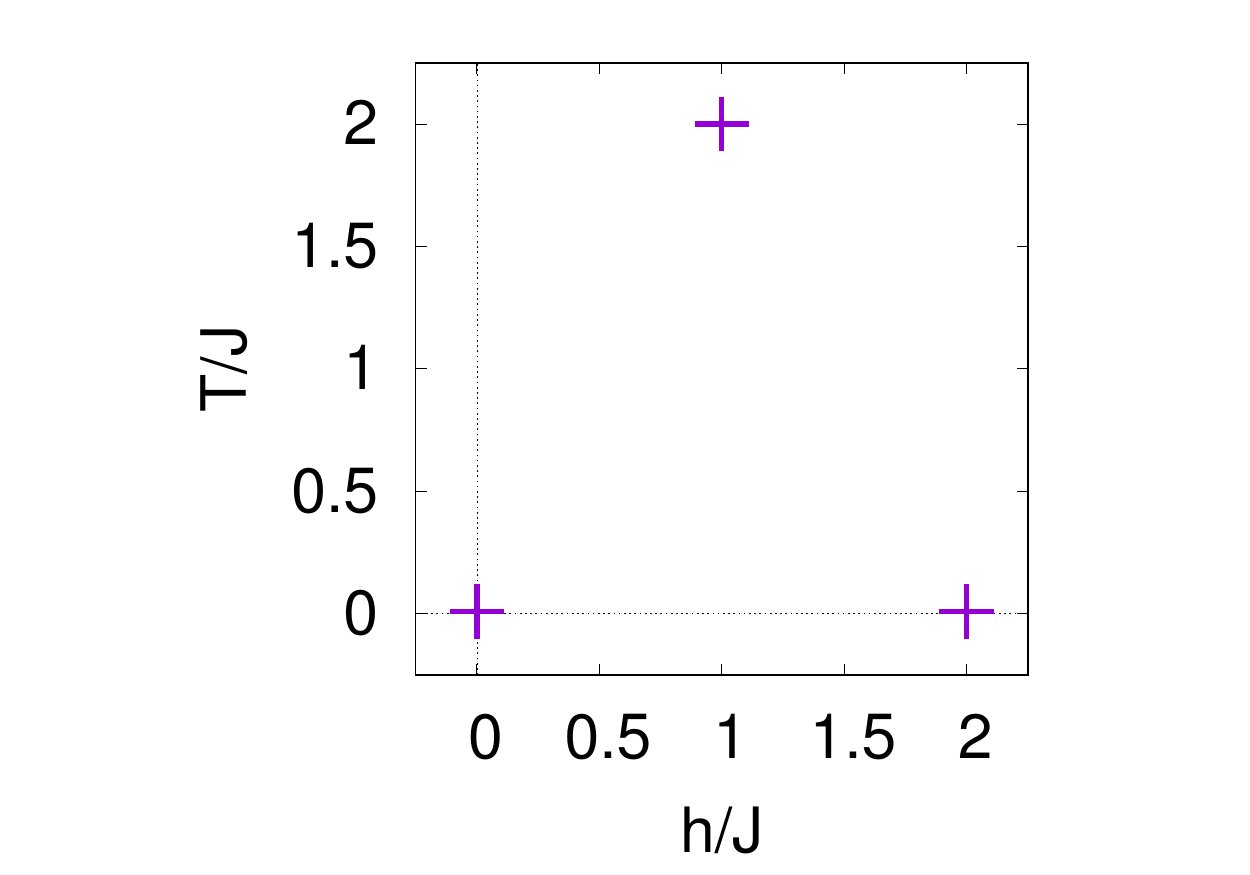}
        \caption{}
    \end{subfigure}
    \begin{subfigure}[b]{0.65\columnwidth}
        \centering
        \hspace{-0.85cm}\includegraphics[width=\columnwidth]{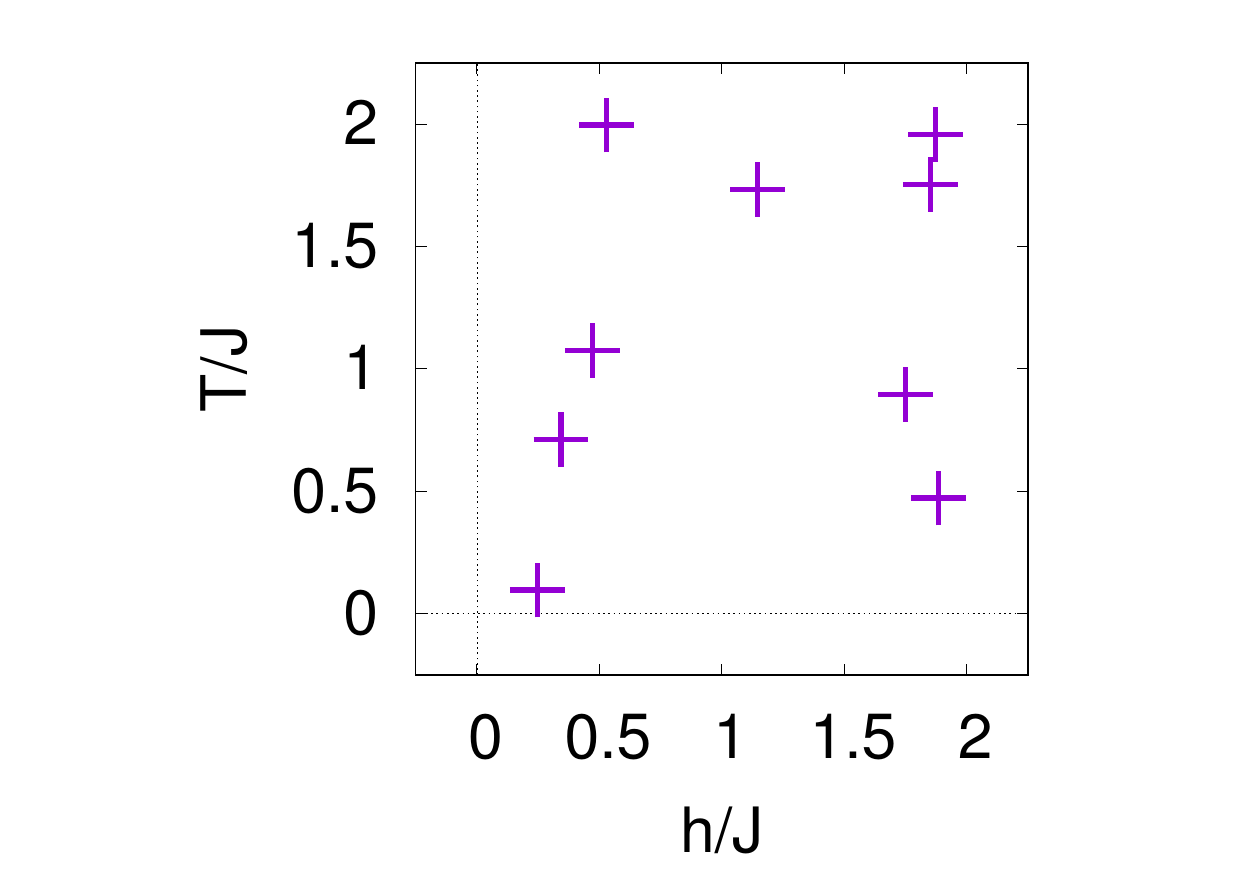}
        \caption{}
    \end{subfigure}
    \begin{subfigure}[b]{0.65\columnwidth}
        \centering
        \hspace{-0.85cm}\includegraphics[width=\columnwidth]{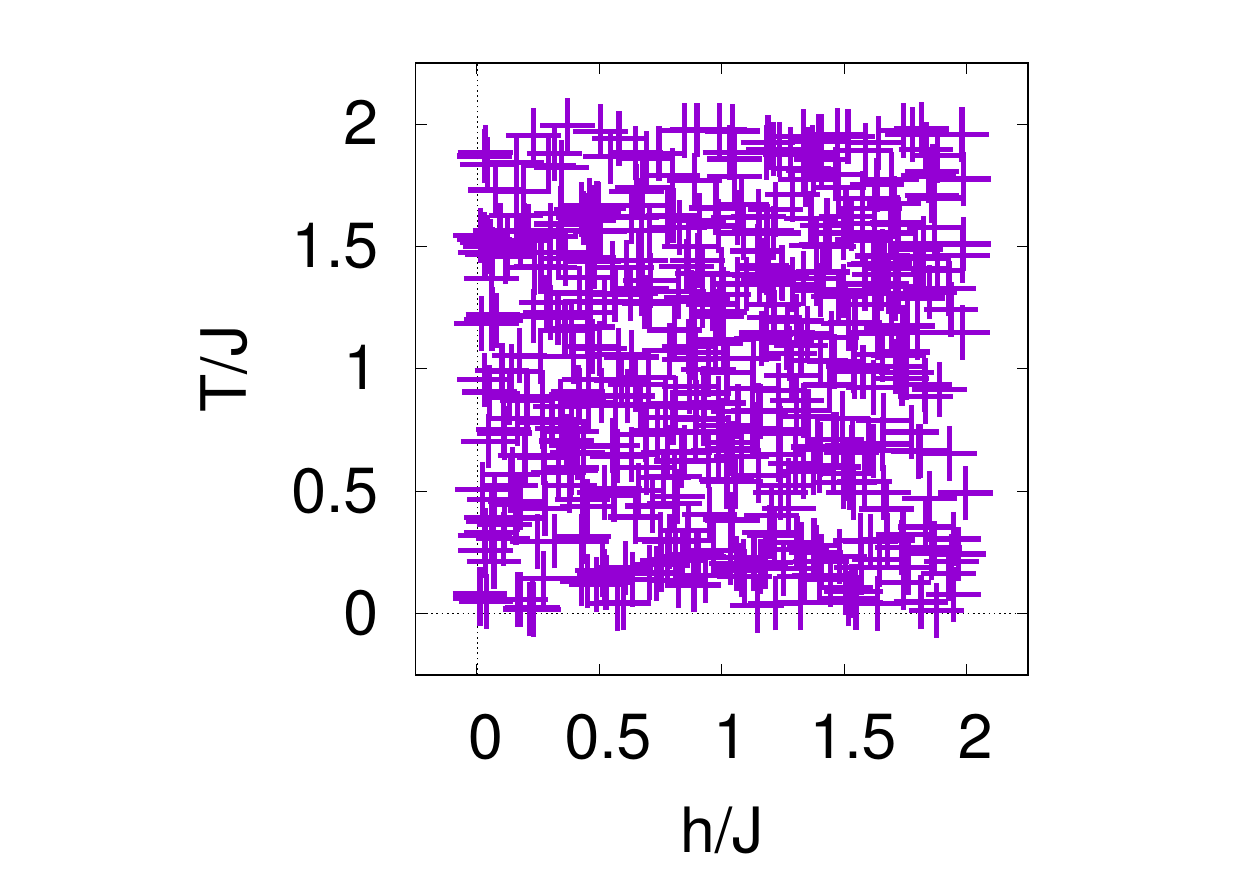}
        \caption{}
    \end{subfigure}
    \begin{subfigure}[b]{0.65\columnwidth}
        \centering
        \hspace{-0.1cm}\includegraphics[width=\columnwidth]{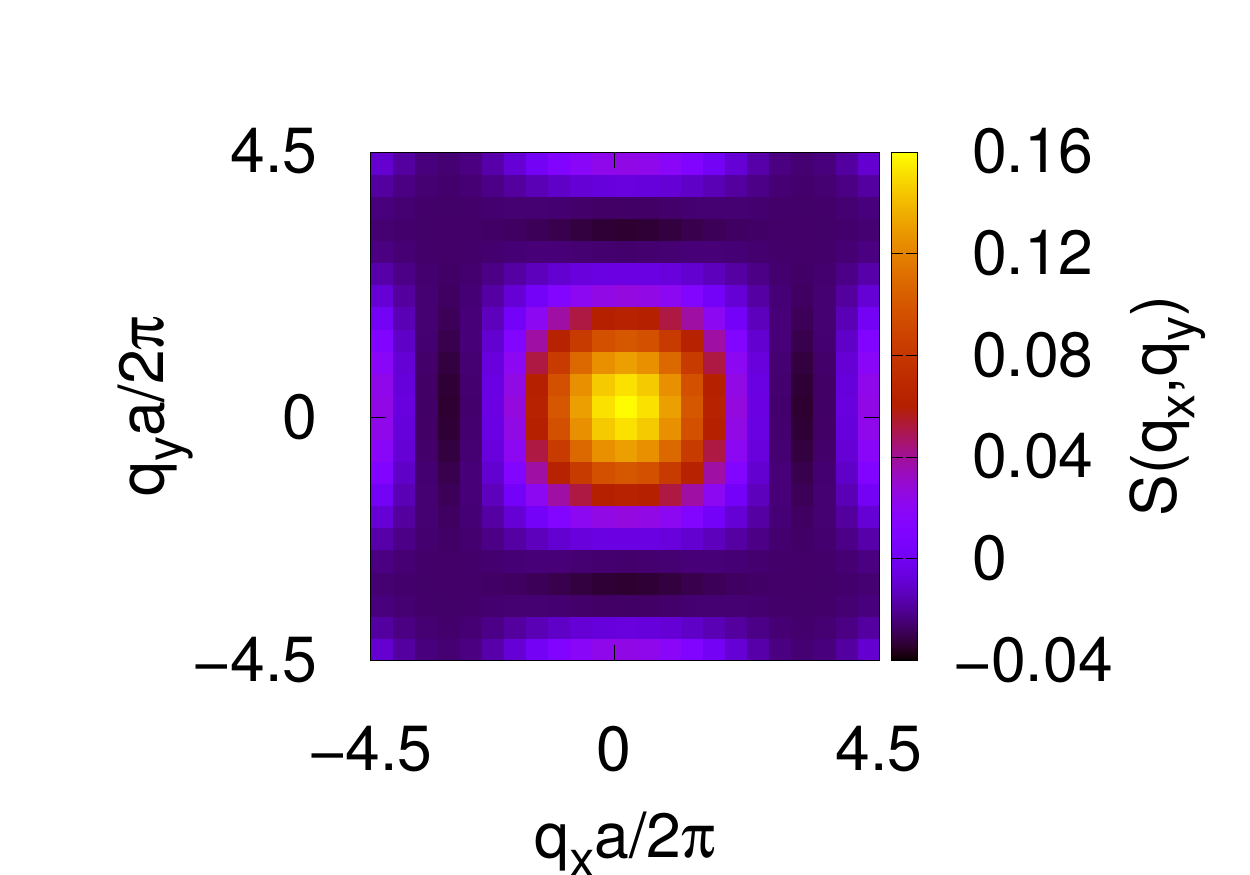}
        \caption{}
    \end{subfigure}
    \begin{subfigure}[b]{0.65\columnwidth}
        \centering
        \hspace{-0.1cm}\includegraphics[width=\columnwidth]{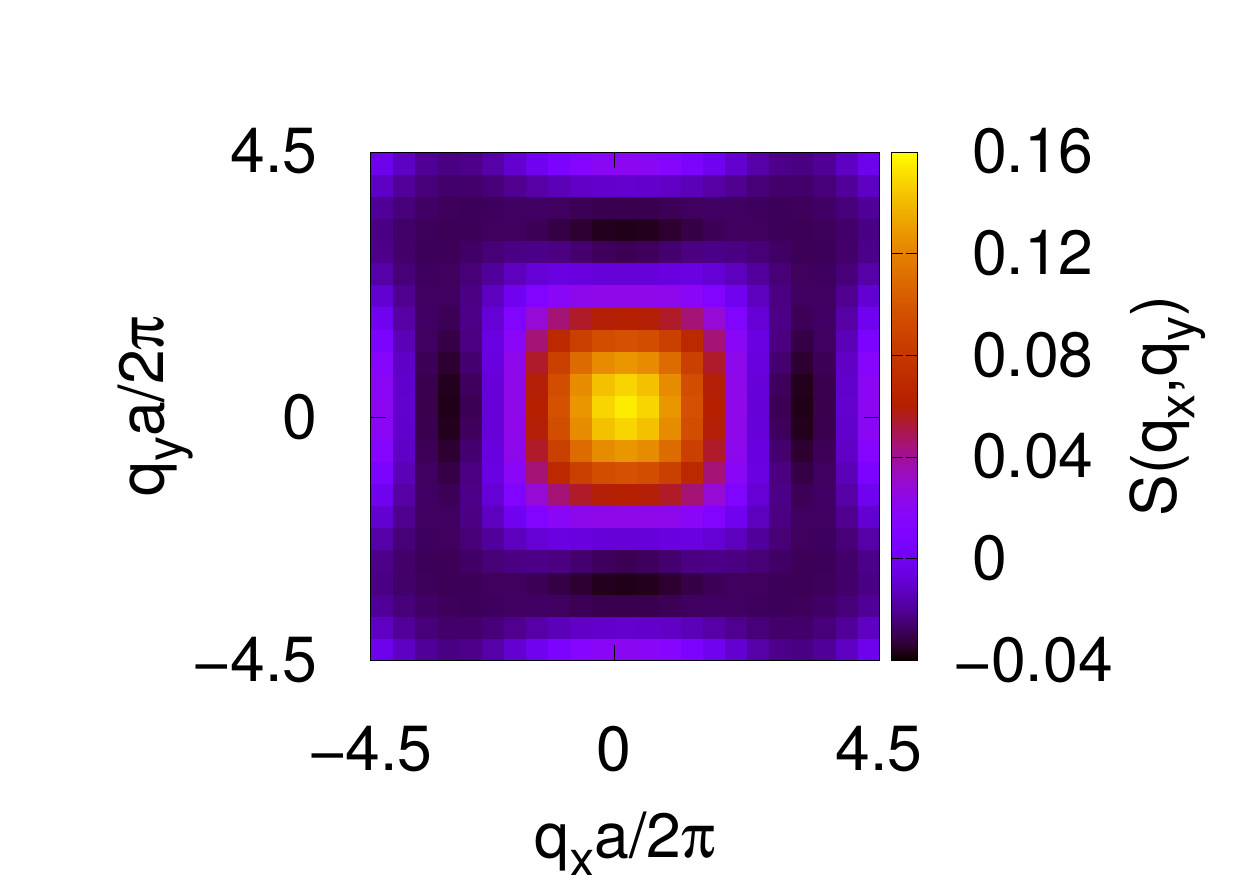}
        \caption{}
    \end{subfigure}
    \begin{subfigure}[b]{0.65\columnwidth}
        \centering
        \hspace{-0.1cm}\includegraphics[width=\columnwidth]{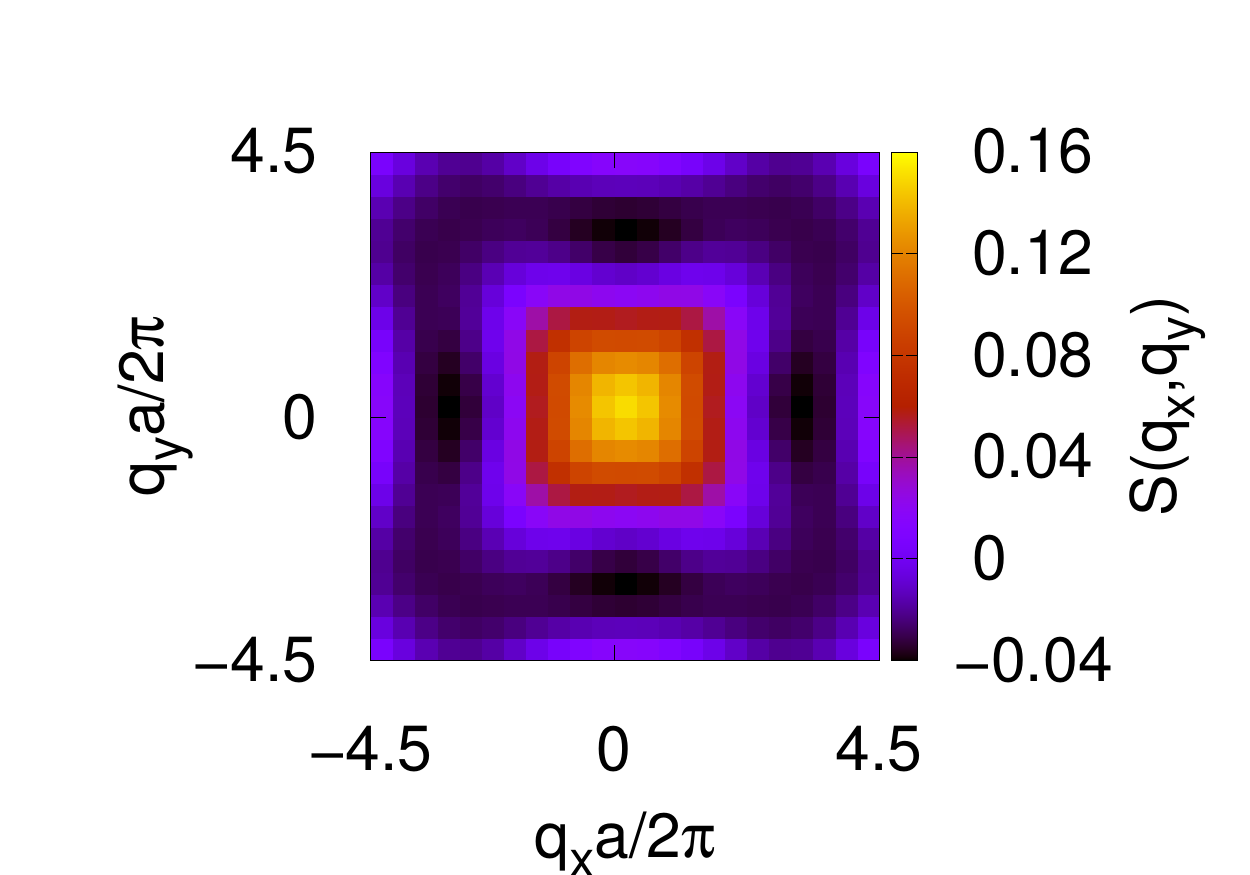}
        \caption{}
    \end{subfigure}
    \begin{subfigure}[b]{0.65\columnwidth}
        \centering
        \hspace{-0.2cm}\includegraphics[width=\columnwidth]{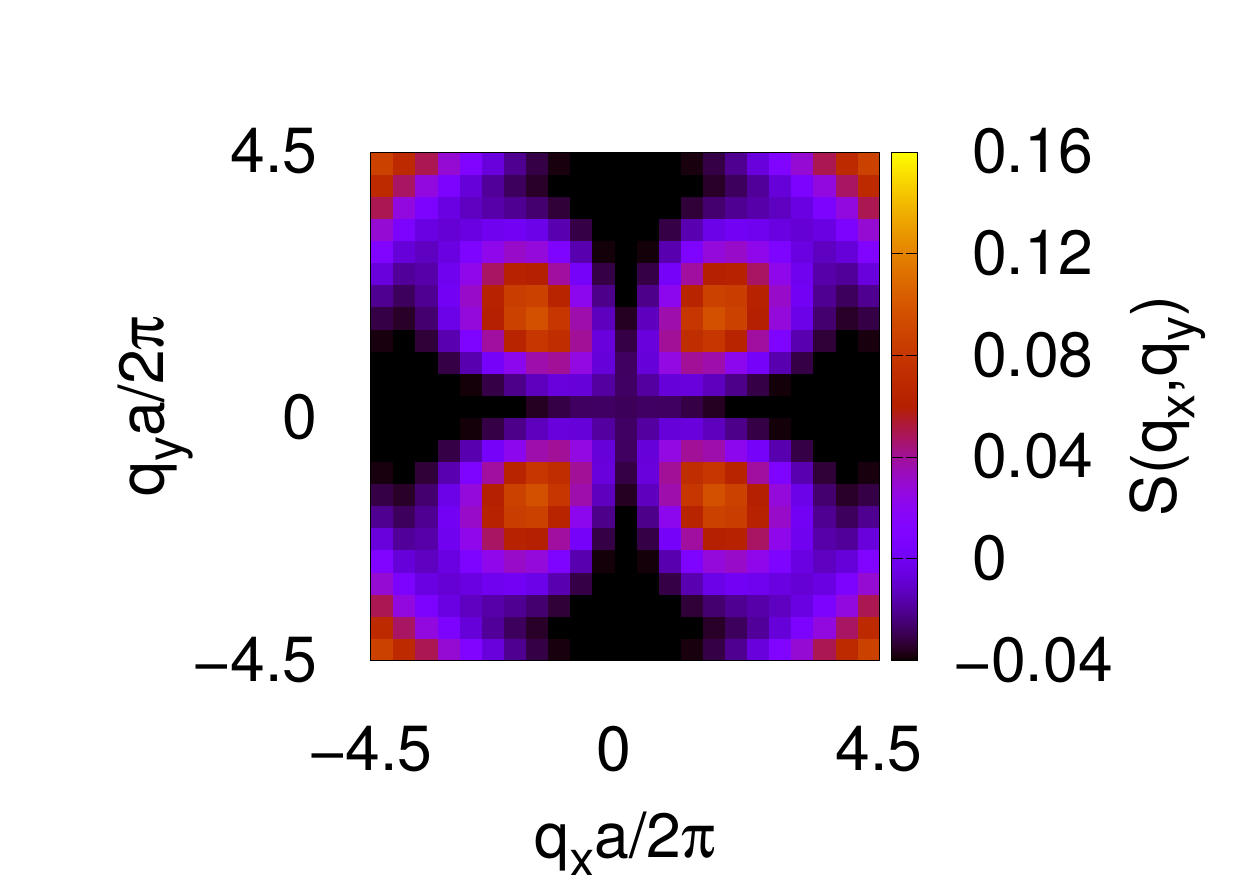}
        \caption{}
    \end{subfigure}
    \begin{subfigure}[b]{0.65\columnwidth}
        \centering
        \hspace{-0.2cm}\includegraphics[width=\columnwidth]{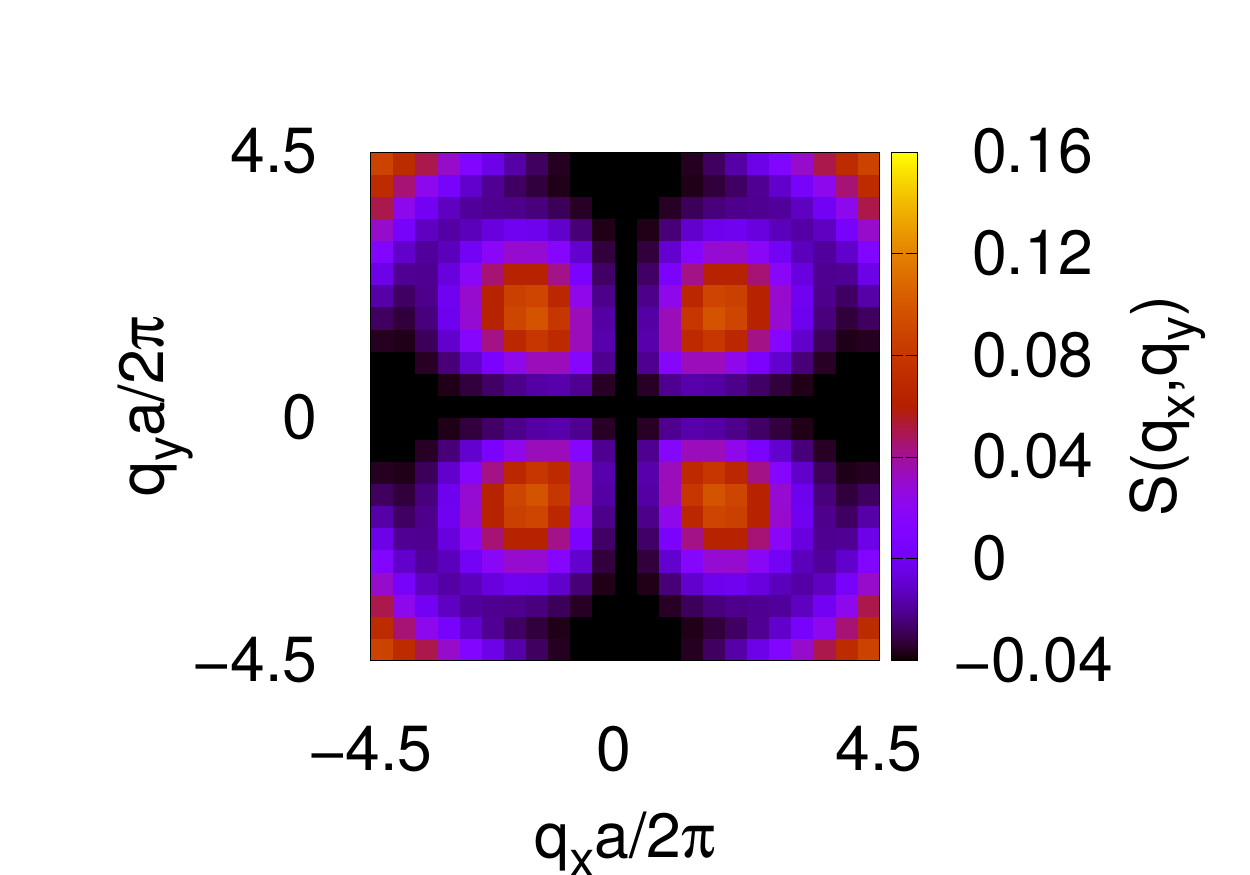}
        \caption{}
    \end{subfigure}
    \begin{subfigure}[b]{0.65\columnwidth}
        \centering
        \hspace{-0.2cm}\includegraphics[width=\columnwidth]{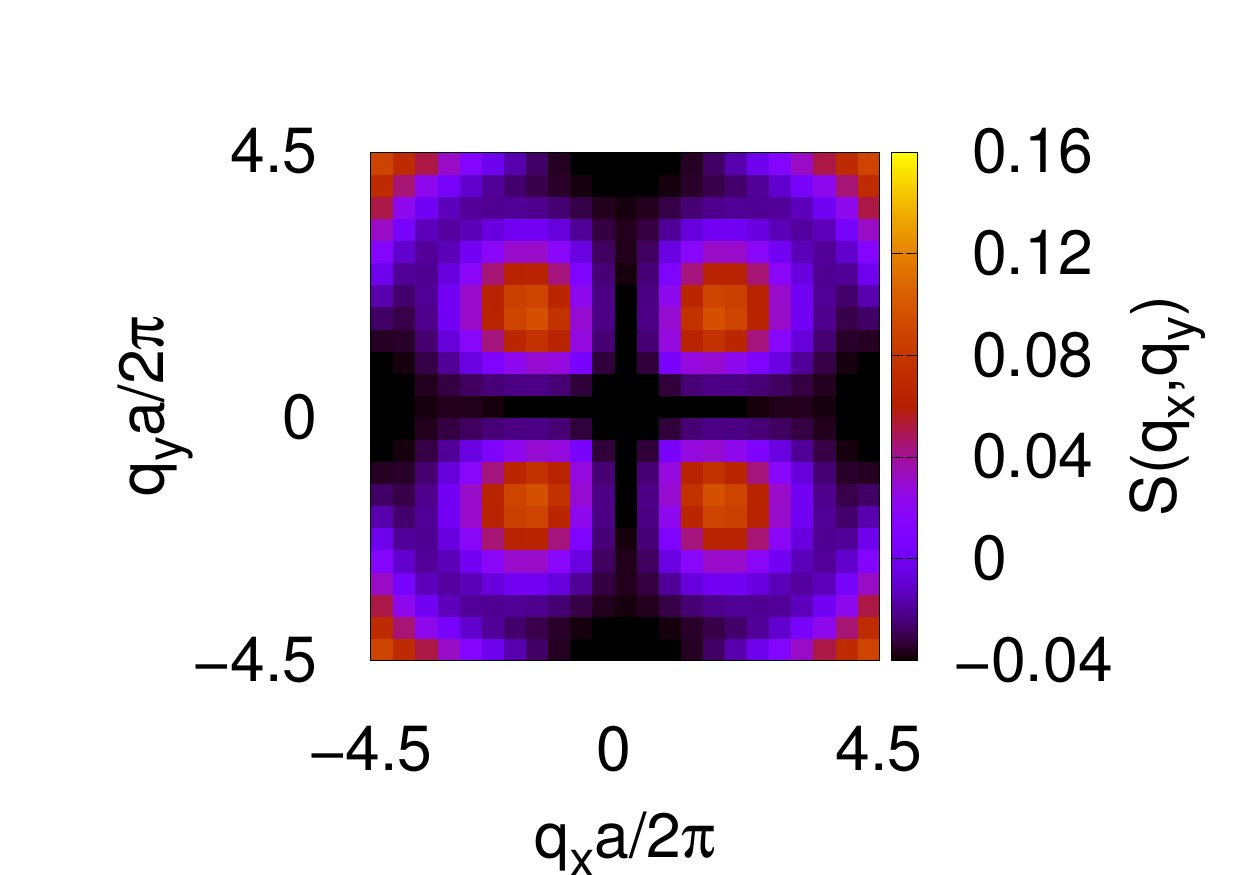}
        \caption{}
    \end{subfigure}

\caption{\label{fig:size_of_training_set}Dependence of the PCs on the choice
of training set for fixed $N=4,\gamma=0.6,\Delta=0$. (a-c) Values
of the magnetic field $h$ and temperature $T$ used to generate each
training set. (d-f) First PC for each of the respective training sets.
(g-i) Second PCs.}
\end{figure*}

\begin{figure*}
\begin{subfigure}[b]{0.5\columnwidth}
    \includegraphics[width=0.95\columnwidth]{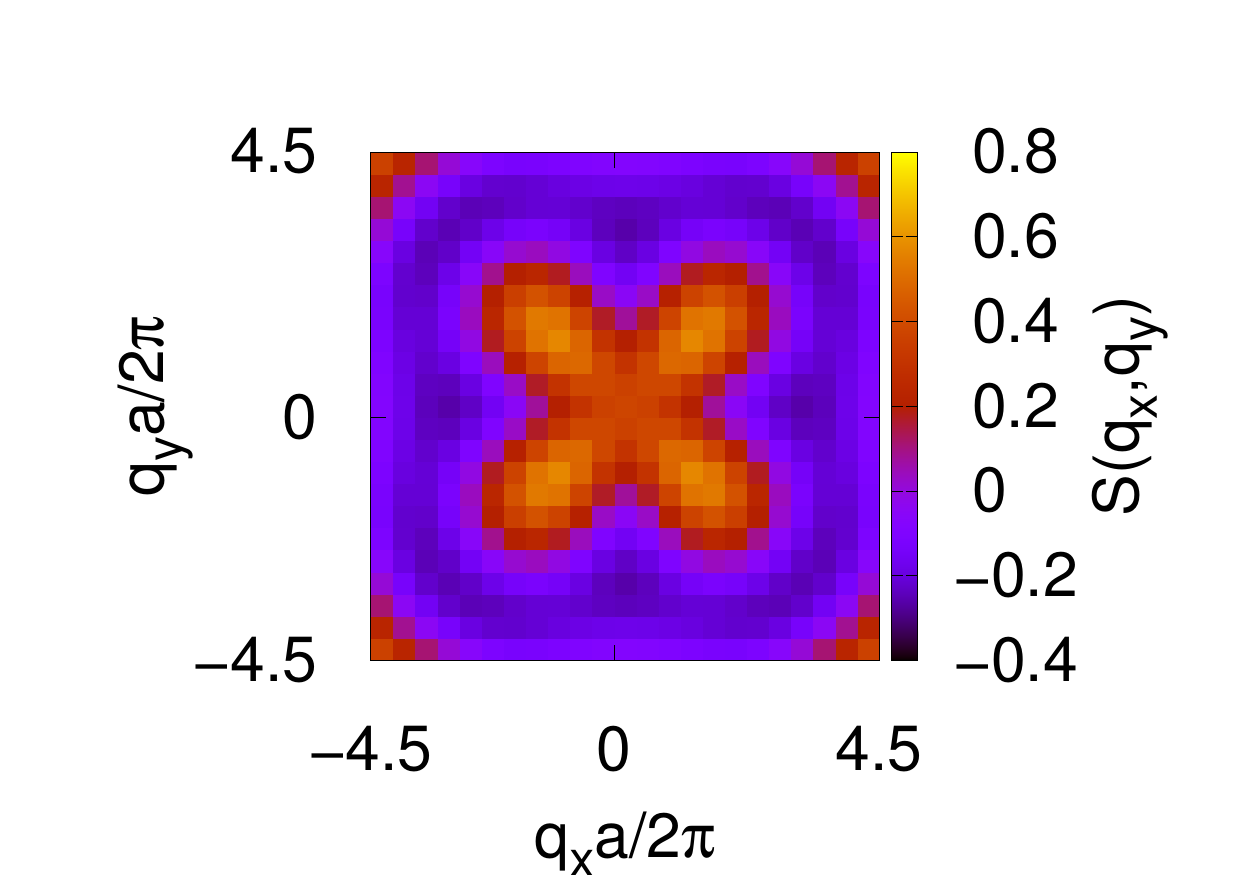}
    \caption{}
\end{subfigure}
\begin{subfigure}[b]{0.5\columnwidth}
    \includegraphics[width=0.95\columnwidth]{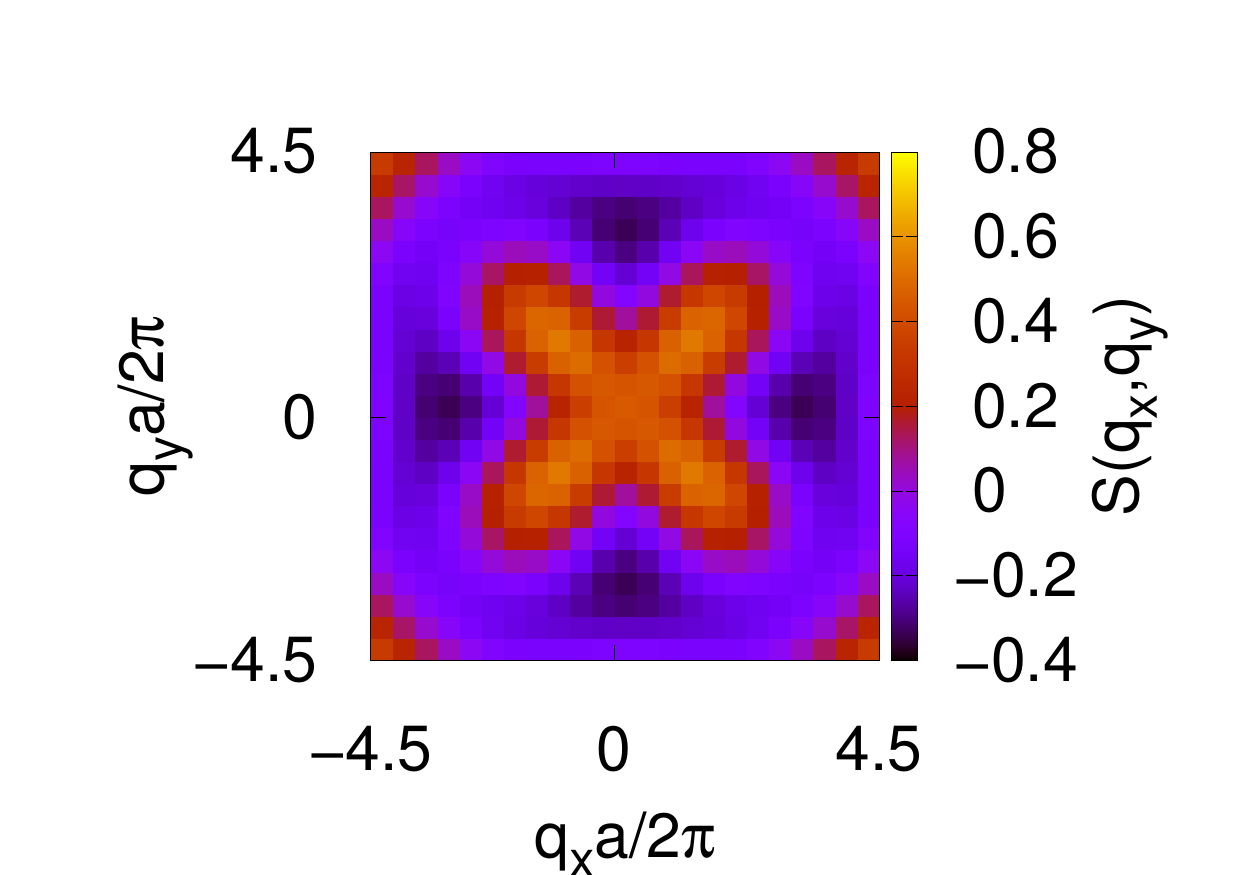}
    \caption{}
\end{subfigure}
\begin{subfigure}[b]{0.5\columnwidth}
    \includegraphics[width=0.95\columnwidth]{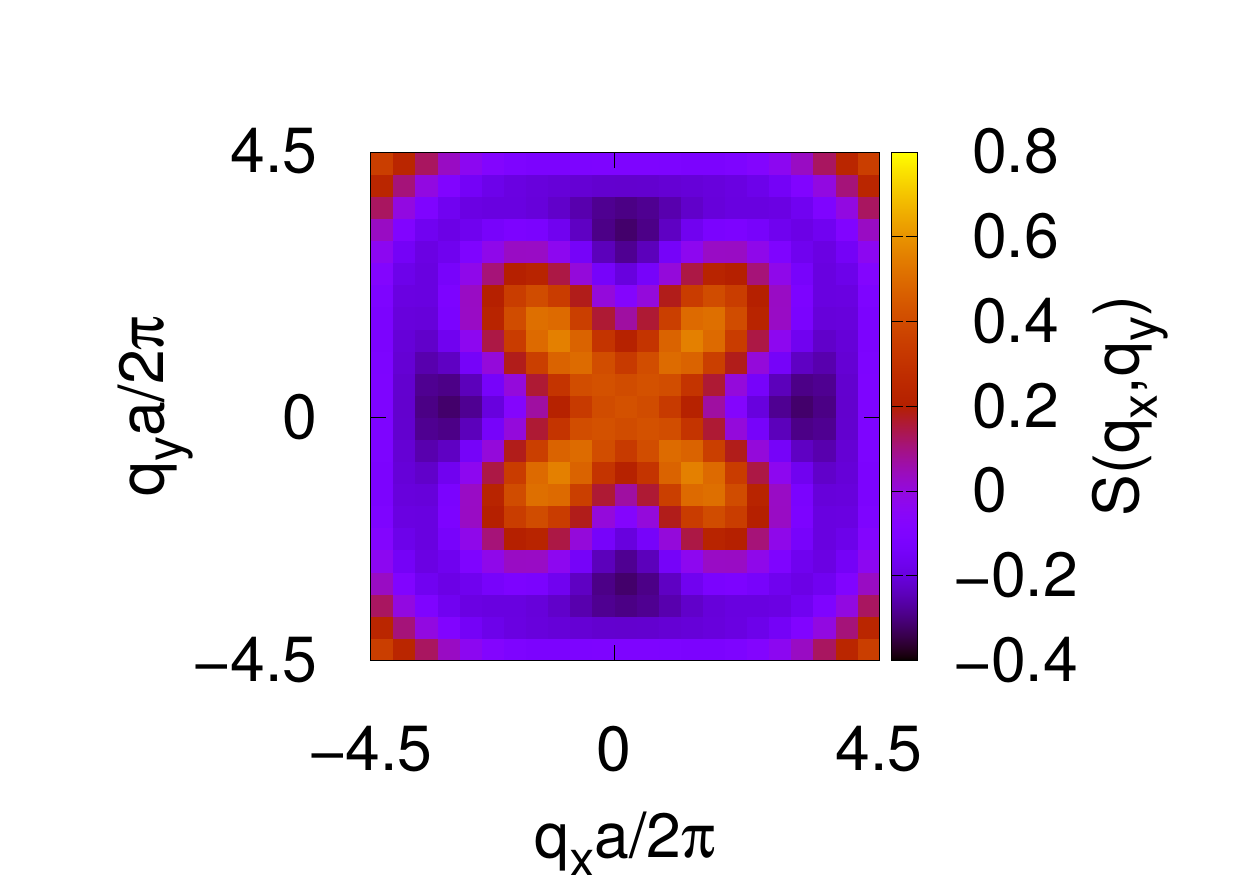}
    \caption{}
\end{subfigure}
\begin{subfigure}[b]{0.5\columnwidth}
    \includegraphics[width=0.95\columnwidth]{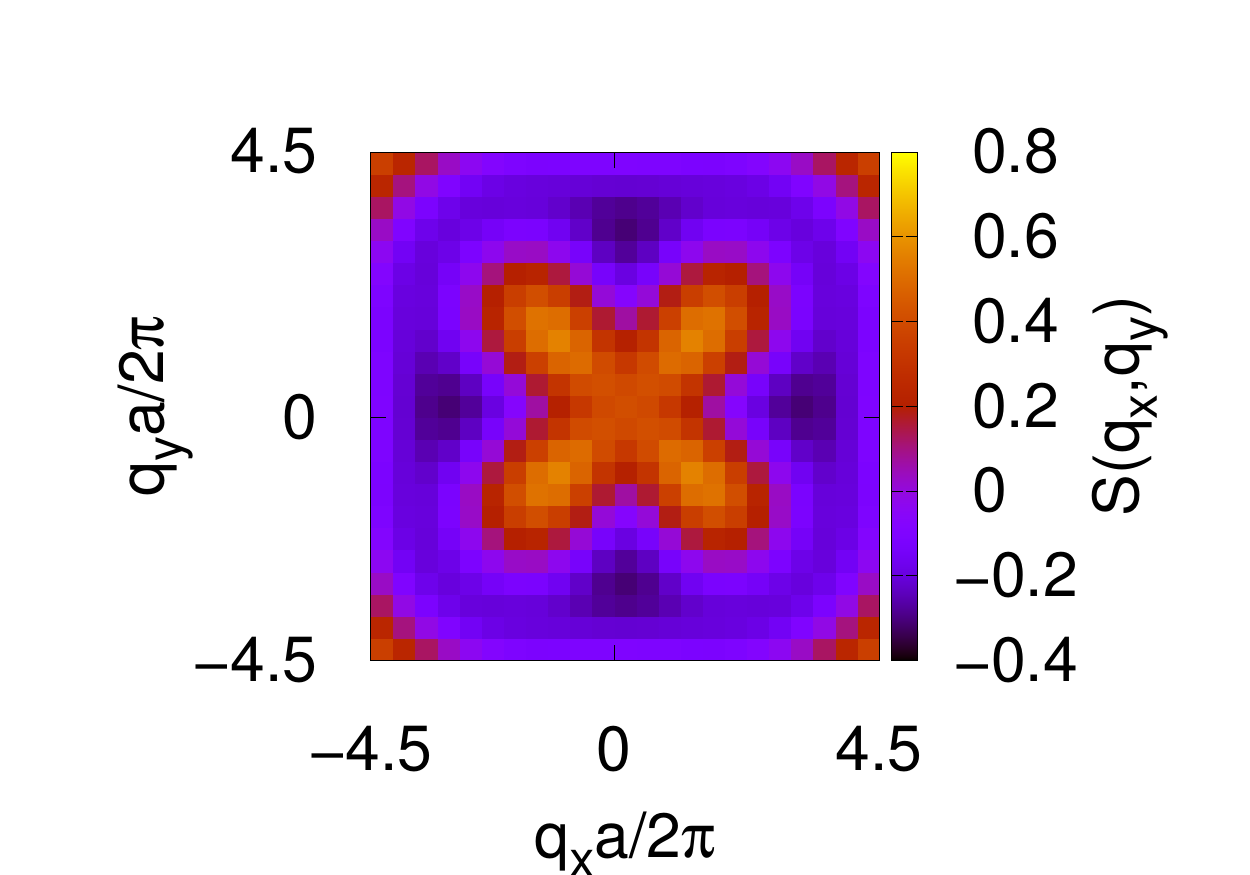}
    \caption{}
\end{subfigure}

\caption{\label{fig:reconstruction}Neutron scattering cross-section $S\left(\mathbf{q}\right)$
predicted by our model for $\gamma=0.6,\Delta=0,h=0.72J,\text{ and }T=0.36J$
(a) and its reconstruction using only two PCs (b-d). The training
sets and their corresponding PCs are the ones shown in Fig.~\ref{fig:size_of_training_set}
consisting of 3 training images (b), 9 training images (c) and 500
training images (d).}
\end{figure*}

Fig.~\ref{fig:size_of_training_set} shows three versions of the
first 2 PCs for a particular instance of our model ($N=4$,$\gamma=0.6$, $\Delta=0$). Each version has
been obtained using a different training set, shown graphically in
the first row of panels: a large training set obtained from 500 randomly-chosen
values of $\left(h,T\right)$; a much smaller training set generated
from 9 randomly-chosen values of $\left(h,T\right)$; and a minimal
set formed by 3 values of $\left(h,T\right)$, chosen strategically
(one with high $T$ one with low $T$ and low $h$, and one with low
$T$ and high $h$) \footnote{Our results are robust with respect to the choice of realisation of a random data set of a given size.}. Appendix \ref{sec:details_of_training_data_set} gives further details about the smaller of the two random sets, including a display of the individual simulated neutron scattering images that compose it.
Our results indicate that $M_{t}$ can indeed
be very small. This is consistent with the small number of principal components necessary to describe the data accurately (see previous section) and is confirmed by Fig.~\ref{fig:reconstruction}
which shows the reconstruction of a particular instance of $S\left(\mathbf{q}\right)$
using the three sets of PCs. We note that the reconstructed image
is not present in any of the training sets.



\section{\label{sec:Bifurcation-plots}Score bifurcation plots}

Figs.~\ref{fig:scree_plots}, \ref{fig:size_of_training_set} and
\ref{fig:reconstruction}, taken together, imply that a relatively
small number of initial observations can be used to determine a few
PCs in terms of which data obtained subsequently can be accurately
described - in other words, by projecting new observations onto the
low-dimensional space spanned by the PCs, we obtain a low-dimensional
representation of the data in terms of the PC scores.
The dependencies of such scores on system parameters such as magnetic
field $h$ or temperature $T$ can then be used to identify features
in the phase diagram. Such approach, when applied to microscopic data
on classical states, has been shown capable of detecting, for example,
symmetry-breaking phase transitions \citep{Hu2017Jun}. Here we ask
whether the same benefit can be obtained when working with observable
averages such as $S\left(\mathbf{q}\right)$ for our quantum magnet
model. We will answer in the affirmative and moreover present a useful
analytical tool based on this idea, which we call a ``score bifurcation
plot''. Our simulations indicate that this technique may facilitate
the detection of qualitative changes in the ground state of a quantum
system, even from data taken at relatively high temperatures.


\begin{figure*}
\begin{subfigure}[b]{0.65\columnwidth}
    \hspace{-0.15cm}\includegraphics[width=\columnwidth]{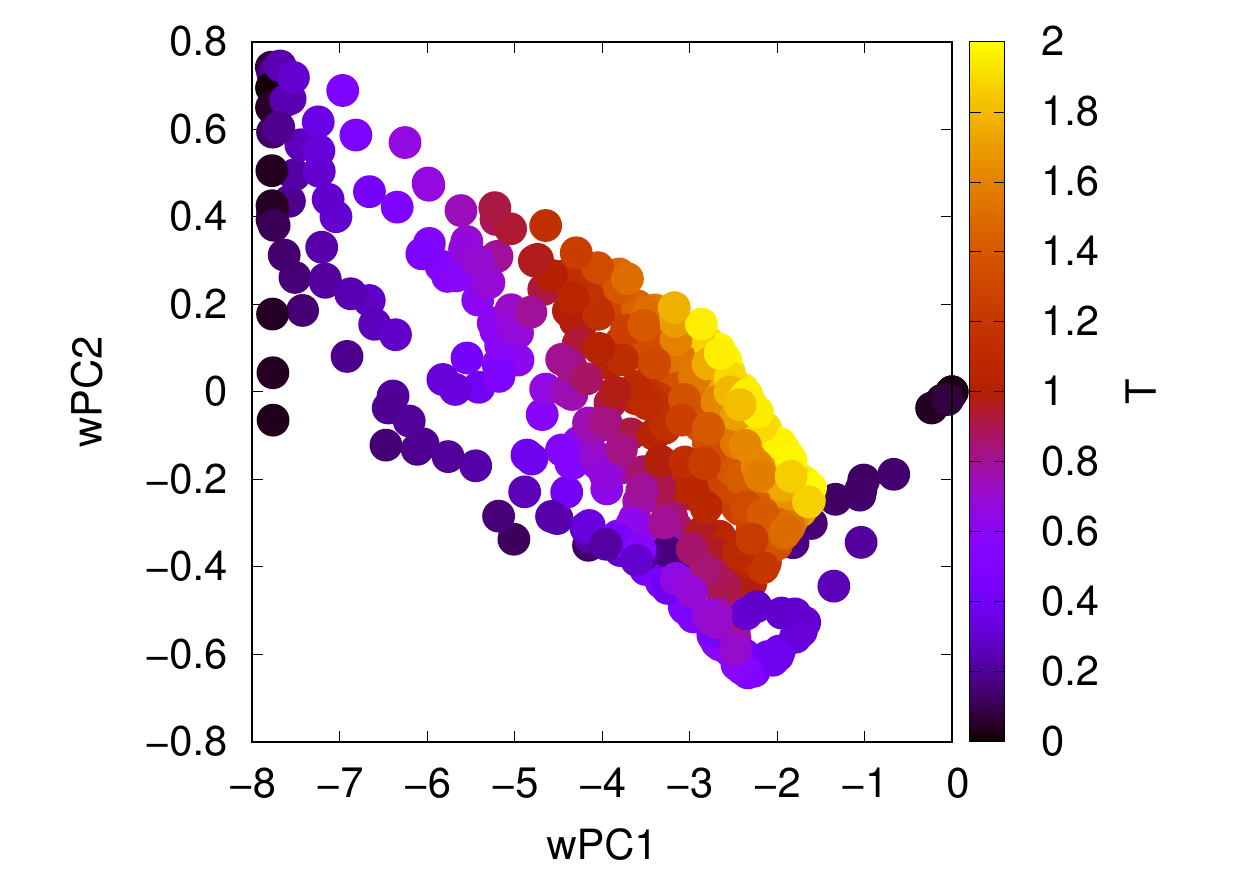}
    \caption{}
\end{subfigure}
\begin{subfigure}[b]{0.65\columnwidth}
    \hspace{-0.15cm}\includegraphics[width=\columnwidth]{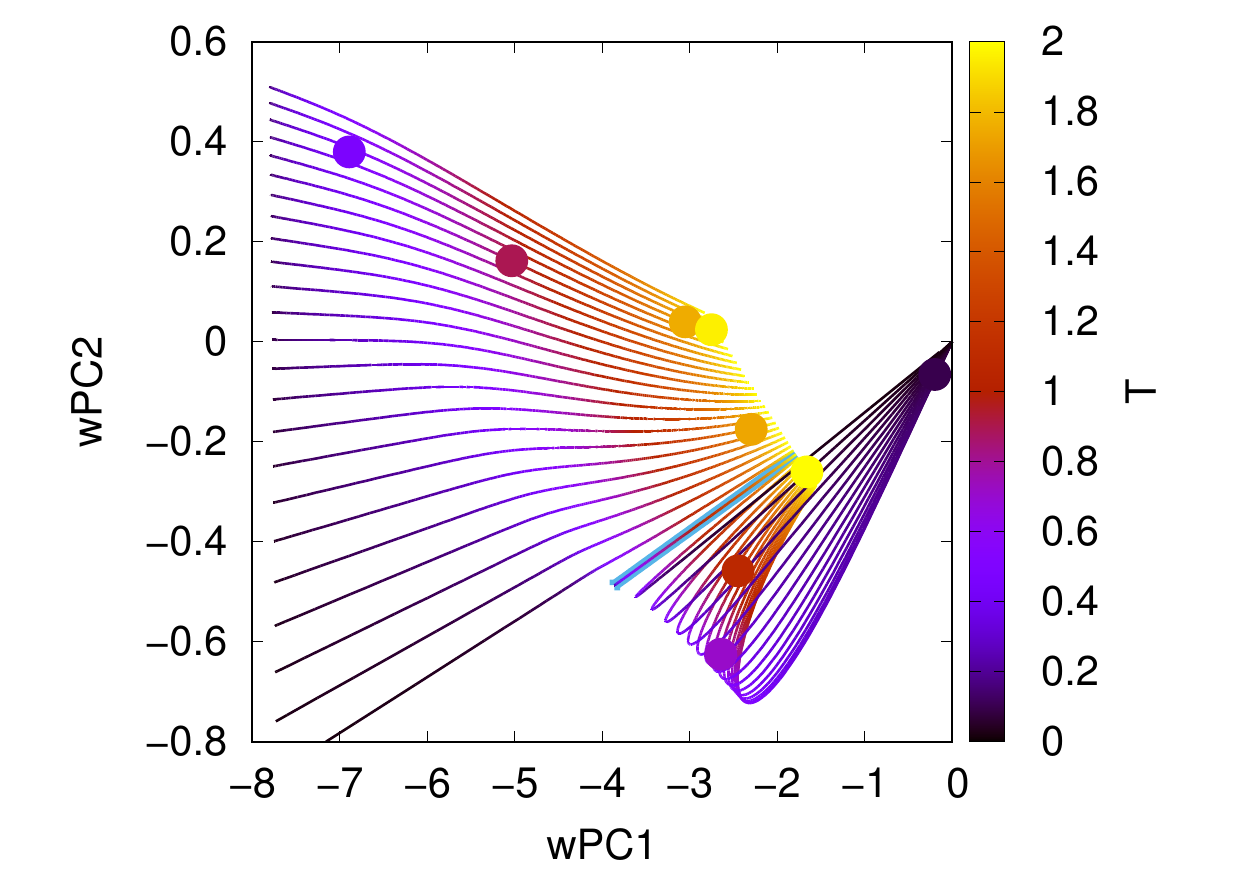}
    \caption{}
\end{subfigure}
\begin{subfigure}[b]{0.65\columnwidth}
    \hspace{-0.15cm}\includegraphics[width=\columnwidth]{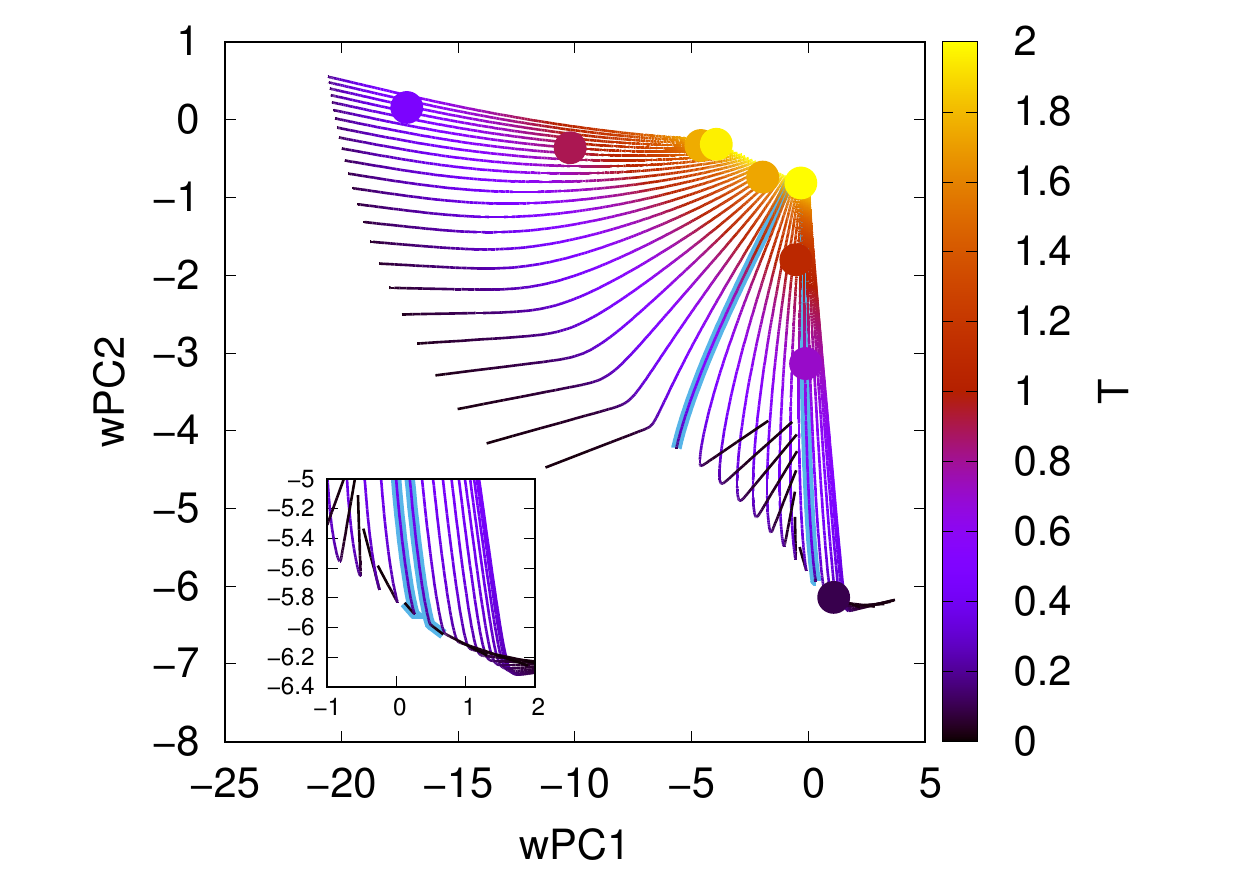}
    \caption{}
\end{subfigure}

\caption{\label{fig:bifurcation_N2}Projections of the simulated scattering
functions $S\left(\mathbf{q}\right)$ of our model for different values
of the field $h$ and temperature $T$ onto the first two PCs. The
interaction parameters $\gamma,\Delta$ are fixed to the same values
as in Figs.~\ref{fig:size_of_training_set},\ref{fig:reconstruction}.
The number of spins in the cluster is $N=2$ (a,b) and $N=4$ (c).
Each scattering image is represented approximately by a single point
(wPC1,wPC2) on the plane defined by the two PCs. The training sets
for panel (a) were obtained using the values of $\left(h,T\right)$
shown in Fig.~\ref{fig:size_of_training_set}~(c); for panels (b,c)
the values shown in Fig.~\ref{fig:size_of_training_set}~(b) were
used. Filled circles represent the scattering functions in the corresponding
training set. Lines in panels (b,c) correspond to additional values
of $\left(h,T\right)$ not present in the training set. These have
been obtained by varying $h$ from 0 to $2J$ in steps of $\Delta h=0.05J$
and $T$ from $2J$ to $0.01J$ in steps of $\Delta T=0.01J$. Each line corresponds to a fixed value of the field $h$. The isolines corresponding to the lowest and highest field values are the the ones reaching furthest to the right and furthest to the top of the graph, respectively. The
curves highlighted in cyan are those for which the field is within
$\Delta h/2$ of the factorisation field $h_{f}=4J/5$ (b,c) or the
additional ground-state level-crossing field $h_{1}=0.35J$ (c). The
inset to panel (c) shows in detail the low-temperature behaviour near
$h\approx h_{1}$. The colour in all panels encodes temperature, as
indicated. Note that the axes limits differ in each case since (a) and (b) are plots obtained using principal components for the same physical system but generated using 9 and 500 images respectively and (c) represents a different system altogether.}
\end{figure*}

To illustrate our method we consider first the simplest case of our
model where $N=2.$ Such $S=1/2$ dimer has two possible ground states:
a low-field anti-ferromagnetic state $|\uparrow\downarrow\rangle-|\downarrow\uparrow\rangle$
with anti-parallel entanglement and a high-field ferromagnetic state
$|\uparrow\uparrow\rangle+\delta|\downarrow\downarrow\rangle$ with
parallel entanglement (as $h\to\infty$ the amplitude $\delta\to0$
resulting in the classical state $|\uparrow\uparrow\rangle$). The
two states are degenerate at the factorisation field $h_{f}$. Panels
(a) and (b) of Fig.~ \ref{fig:bifurcation_N2} show the weights,
or scores, of the first two principal components, wPC1,wPC2, obtained
by projecting the PCs onto the simulated neutron scattering function
$S\left(\mathbf{q}\right)$, for two different training sets. In panel
(a) the weights have been obtained for the scattering functions in
the original training set. We can appreciate a marked difference between
the high-temperature data, concentrated in a small region of wPC1-wPC2 
space, and the low-temperature data which cover a much wider area.
This is reminiscent of results obtained for microstates of Ising-type
models ---c.f. Fig.~3~(b) in Ref.~\citep{Hu2017Jun}, where we also see that, at high temperatures, the data are clustered in a small, high-entropy region of PC space while, at lower temperatures, the data fan out as distinct ground states are selected by their energies. In our case,
however, we are examining the statistical average $S\left(\mathbf{q}\right)$
which is a function of parameters $h,T$ that can, in an experimental
situation, be controlled externally. It is therefore possible to explore
the space of scattering functions systematically by varying $h$ and
$T$ and projecting the new measurements onto the PCs deduced from the training set. A simulation of that approach is presented in panel
(b), which shows the evolution of wPC1 and wPC2 with temperature for
different fixed values of the magnetic field (fixed-field temperature
scans). We observe a marked difference between the curves corresponding
to fields lower than the factorisation field $h_{f}$ and those greater
than that field. Each constant-field temperature scan is represented
by a single curve in the space defined by the two principal components.
As the temperature is lowered, the curve starts to bend in a direction
that indicates the nature of the ground state (ferromagnetic if $h>h_{f}$
and anti-ferromagnetic if $h<h_{f}$). This manifests as a marked
bifurcation, with the curve corresponding to $h=h_{f}$ (highlighted
in cyan) marking the bifurcation where the directions of this bending
changes sign. At this field, below some finite temperature $T^{*}$
the system gets ``stuck'' at a particular point (wPC1$^{*}$,wPC2$^{*}$)
and does not evolve further. This suggests that such ``score bifurcation
plots'' can be used to elucidate systematically the ground-state
phase diagram from finite-temperature data, even in systems such as
the one we model where there are no finite-temperature phase transitions. Appendix \ref{sec:geometric_construction} describes a geometric construction that can be used to determine $h_f$ accurately from limited finite-temperature data.

Further evidence of the above hypothesis is provided in Fig.~\ref{fig:bifurcation_N2}~(c)
where similar data are presented for $N=4$. As we reviewed in Sec.~\ref{sec:Model},
two special values of the field $h_{1},h_{f}$ are expected to emerge
at low temperatures in this case. Indeed we find two bifurcations
occurring at those fields, within the resolution given by our field
increment $\Delta h=0.05J$. We note however that the bifurcation
at $h_{f}$, where the ground state factorises leading to the vanishing
of entanglement measures, is detectable at a higher temperature than
that at the level crossing field, where entanglement is suppressed
but does not vanish \citep{Irons2017}.
We have verified that the factorisation field is also seen in similar
score bifurcation plots obtained for $N=6$.

\section{\label{sec:Experimental-resolution}Experimental resolution}

In Sec.~\ref{sec:Dimensionality-reduction} we showed that PCA could
achieve good dimensionality reduction for neutron scattering data
simulated using our model. In Sec.~\ref{sec:Size-of-training} we
further showed that this could be achieved using surprisingly small training sets, in the
sense of containing very few observations. We will now address the question of experimental resolution
- specifically, how many pixels each individual observation needs
to have for the score bifurcation plots introduced in the last section to
yield accurate values of the critical fields.
We note that modern neutron scattering instruments
allow a trade off between neutron flux and resolution at the time when the measurement
is being made \citep{Chapon2011}. In addition, one can always
group pixels together, post-measurement, to form a more coarse-grained,
but less noisy image.
We note that the authors of Ref.~\onlinecite{Samarakoon2019} have addressed the question of noise in a different way, namely by applying their methodology directly to noisy images, and their conclusions are similar to ours.

\begin{figure}
\begin{subfigure}[b]{1\columnwidth}
    \centering
    \includegraphics[width=0.95\columnwidth]{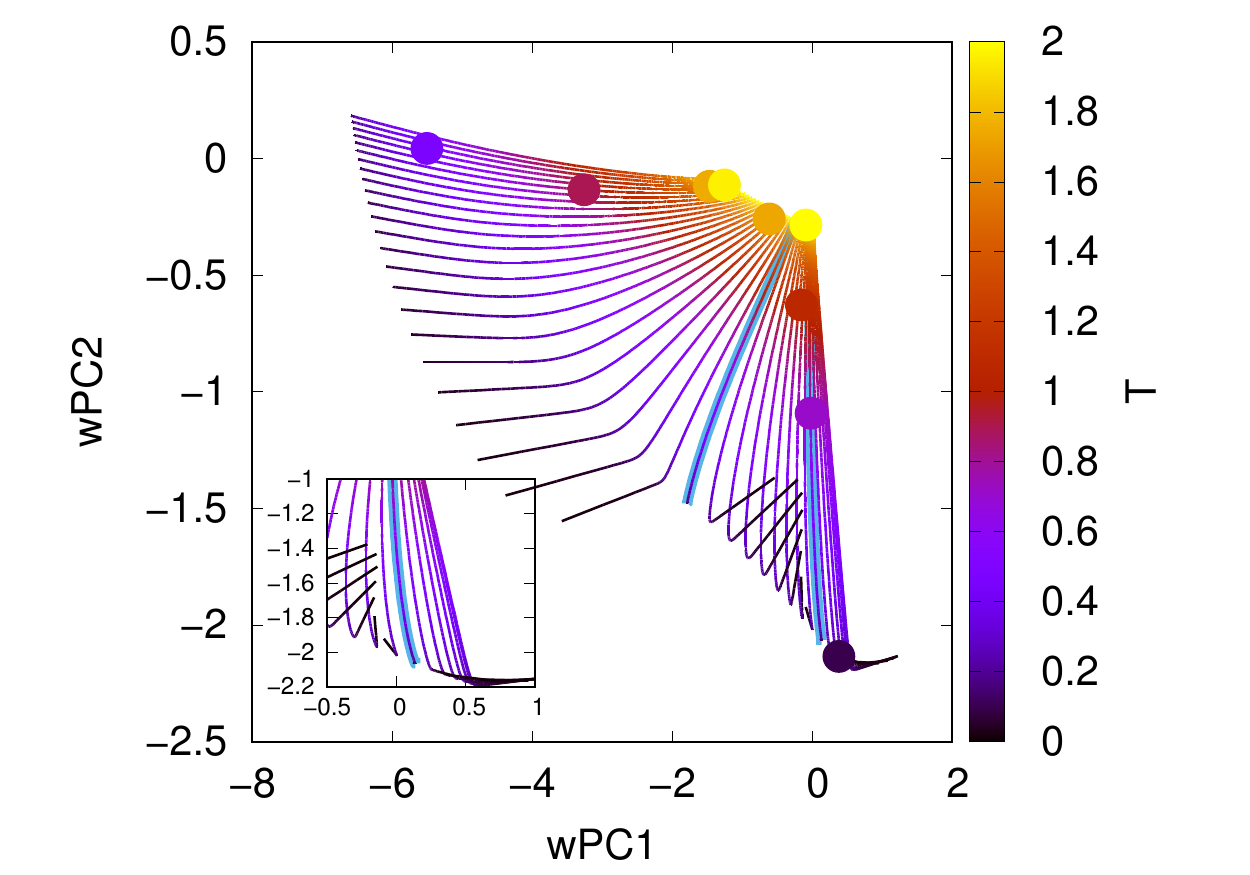}
 \caption{}
\end{subfigure}
\begin{subfigure}[b]{1\columnwidth}
    \centering
     \includegraphics[width=0.75\columnwidth]{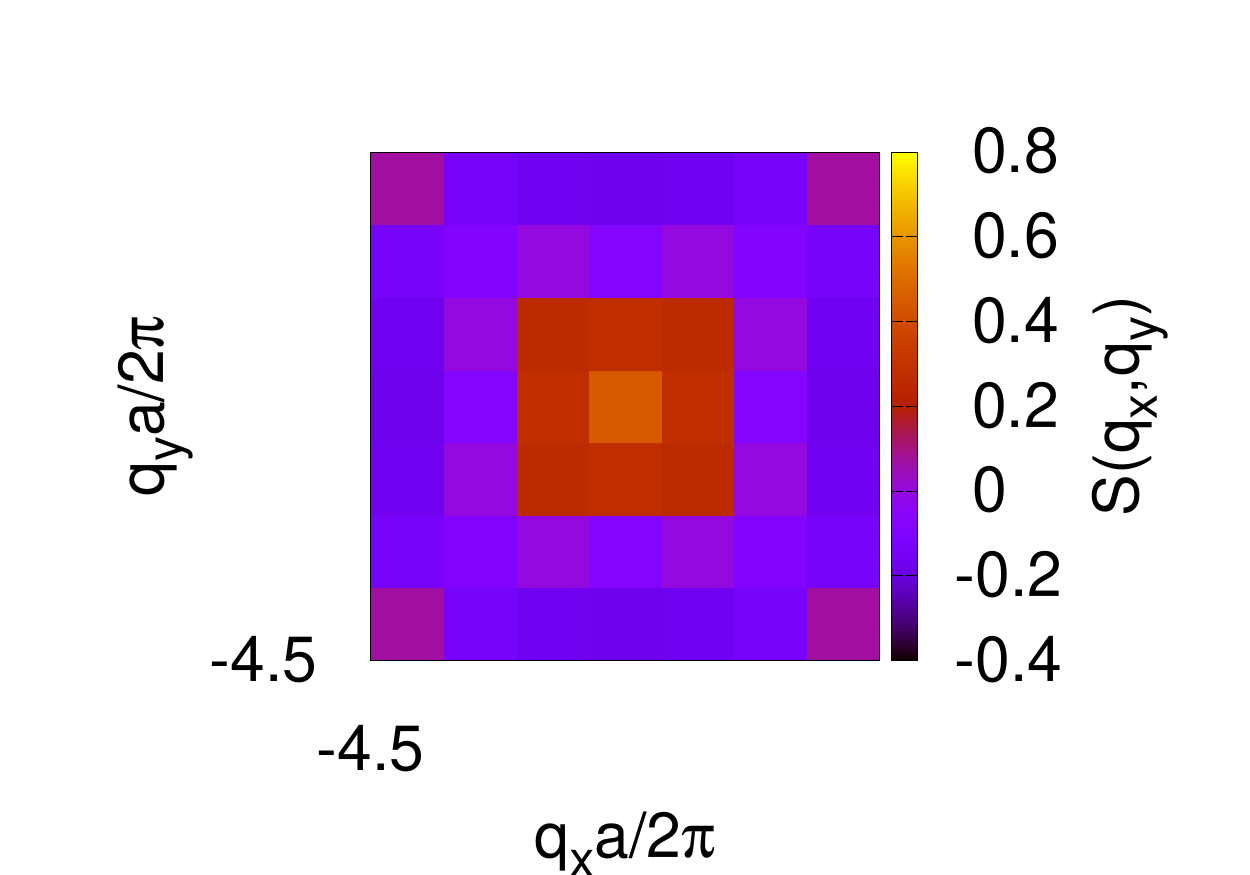}
 \caption{}
\end{subfigure}

\caption{\label{fig:bifurcation_N4_lower_res}(a) Score bifurcation plot for the same
parameters as in Fig.~\ref{fig:bifurcation_N2}~(c) except that
here all observations (both those used in training as well as all
subsequnet observations) consist of much lower-resolution, namely $8\times8$
pixel images. (b) An example of a scattering function obtained by exact diagonalisation at that resolution (the other parameters are as in Fig.~\ref{fig:reconstruction}).}
\end{figure}

To address our question we have repeated some of our previous calculations
using lower-resolution images both at the training and analysis stages.
Specifically, we replace our previous $24\times24$ pixel matrices
with $8\times8$ matrices which corresponds to nearly an order of
magnitude reduction in the amount of data in each observation. The
results, for the $N=4$ case, are shown in Fig.~\ref{fig:bifurcation_N4_lower_res}.
Clearly, the score bifurcation plot obtained from these lower-resolution
images is as useful as that obtained before, and in particular it
allows us to pinpoint the critical fields $h_{1}$ and $h_{f}$ to
the same values, within our field-scan accuracy~$\Delta h=\pm0.025J.$

\section{Discussion and Conclusions}

In this work we have presented a method to obtain quantitative information
about the phase diagram of a quantum system from experimentally observable
data, namely the diffuse magnetic neutron scattering function $S\left(\mathbf{q}\right)$.
Our method is based on a simple form of unsupervised
machine learning, PCA, and provides a visual representation of the data that facilitates a greater understanding of the underlying physics. We addressed our research question theoretically by
analysing simulated scattering functions for a simple model of a cluster
quantum magnet.

Our method is based on using a small training set
of observations to determine PCs describing the data. We then analyse
subsequent observations by projection onto those PCs. We found that
this procedure can achieve a large degree of dimensionality reduction
i.e. very few PCs suffice for an accurate description of subsequent
observations. Consequently, we found that effective
training requires only a small number of observations. Moreover, each
observation may be a very low-resolution image, which should facilitate
greatly the implementation of our methodology in a real experimental
setting. Finally, we devised a way to use the trained PCA algorithm
to characterise the system's phase diagram.

Our method to investigate phase diagrams relies on temperature scans at fixed values of another control parameter
(in our case, magnetic field $h$ - however the method can be trivially
generalised to other control parameters such as pressure). By plotting
the evolution of the system in PC space we find paths with bifurcations
occurring at the values of the field that are known to correspond
to changes in the system's ground state.

It is interesting to speculate why a linear technique, PCA, can provide such a compact and enlightening description of a system dominated by strong correlations. In this respect we note that our calculation of the scattering function $S(\textbf{q})$ was carried out in the linear-response regime~\cite{Irons2017,magneto2020}. This is standard in the theory of magnetic neutron scattering and is justified by the fact that the neutron acts as a weak perturbation~\cite{Lovesey1987b}. Whether this can be used as the starting point for a justification of the applicability of PCA is an interesting question but lies outside the scope of the present work.

We hasten to add that the linear dependence of $S(\mathbf{q})$ on the principal components does \emph{not} imply that $S(\mathbf{q})$ is a linear function of the model parameters $\gamma,\Delta,T,h$. The scattering function is given by
\begin{equation}
    S(\mathbf{q}) = \sum_n w_n(\gamma,\Delta,h,T) S_n(\mathbf{q})
    \label{eq:exp}
\end{equation}
where the score $w_n(\gamma,\Delta,h,T)$ of the $n^{\mbox{\underline{th}}}$ PC is in general a highly non-linear function of $\gamma,\Delta,h$ and $T$. Thus in our dimensionally-compressed representation of the data $S(\mathbf{q})$ is a linear function of just a few $S_n(\mathbf{q})$'s but the non-linearity of the model is still reflected in the dependence on system parameters of the expansion coefficients. It is nevertheless remarkable that the non-linear features of the model can be accurately captured in this way. An explicit example of such non-linear behaviour of the PC scores is  provided in Appendix~\ref{sec:individual_scores}.

Our method provides an efficient
way to extrapolate the ground-state phase diagram from finite-temperature
data, even in a system that is effectively of finite size and therefore
lacks well-defined, finite-temperature phase boundaries. One may speculate that applying
such methodology to poorly-understood real systems such as copper-based
high-temperature superconductors might offer a fresh perspective on
an old conundrum in the theory of strongly-correlated electron systems:
are the correlated quantum ``liquid'' phases found at finite temperature best thought of as
manifestations of the quantum critical points separating the different
quantum-ordered phases (as proposed by Laughlin and co-workers \citep{Laughlin2010Nov})?
Or are they best regarded, instead, as condensations of the ``gaseous'' phase existing at higher
temperatures, which become susceptible to different quantum ordering
transitions as the temperature is lowered further (as put forward
by Anderson \citep{Anderson2002})? Addressing this question will require applying our method to experimental data on real systems. For instance, a score bifurcation plot of measurements taken in the ``strange metal'' phase of the cuprate phase diagram may provide evidence for proposals that puzzling crossovers observed in that region are due to a Widom line associated with a hidden, low-temperature critical endpoint~\cite{sordi_pseudogap_2012}.

\appendix

\section{PCA Implementation}


The PCA was implemented in the Octave (v3.4.3) programming language~\cite{OctaveCite} using the singular value decomposition function which returns singular vectors normalized to unit length. A is a matrix, in which each column is a scattering function image represented in column-wise concatenated form. The data is centred by subtracting from each column its mean pixel intensity value, thereby forming matrix X.  This is followed singular value decomposition from which PCs, PC scores, and scree plots are straight-forward to obtain. We reproduce the key part of our code here:\tabularnewline

1:\textbf{\% Data centering:}

2: \textbf{X=A-ones(size(A)(1),1)*mean(A)}

3: \textbf{\% Singular value decomposition:}

4: \textbf{{[}V,lambda,junk{]} = svd(X'{*}X);}

5: \textbf{\% Principal components:}

6: \textbf{U = X{*}V{*}lambda\textasciicircum (-.5);}


7: \textbf{\% Principal component scores:}

8: \textbf{S = U'{*}X;}

9: \textbf{\% Data for scree plot:}

10: \textbf{scree\_data=diag(lambda);}\tabularnewline\tabularnewline
Adapt the code for Matlab by replacing line 2 with \textbf{X=A-ones(size(A,1),1)'{*}mean(A);}

The complete code is available in Ref.~\cite{jarvis2020} as a  Github repository. The codes used to generate the matrix A can be found in Ref.~\cite{magneto2020}.

We note that our re-centering procedure does not involve substracting an average over observations, as is common in other PCA implementations. This essentially means that the first PC describes that average, and the corresponding score quantifies how much a specific observation deviates from it. Our tests indicate that, for the type of data studied here, this yields a clearer and more complete description of the underlying correlations.


\section{\label{sec:details_of_training_data_set}Details of training data set}

\begin{figure*}
    \centering
    \begin{subfigure}[b]{0.65\columnwidth}
        \centering
        \hspace{-0.0cm}\includegraphics[width=\columnwidth]{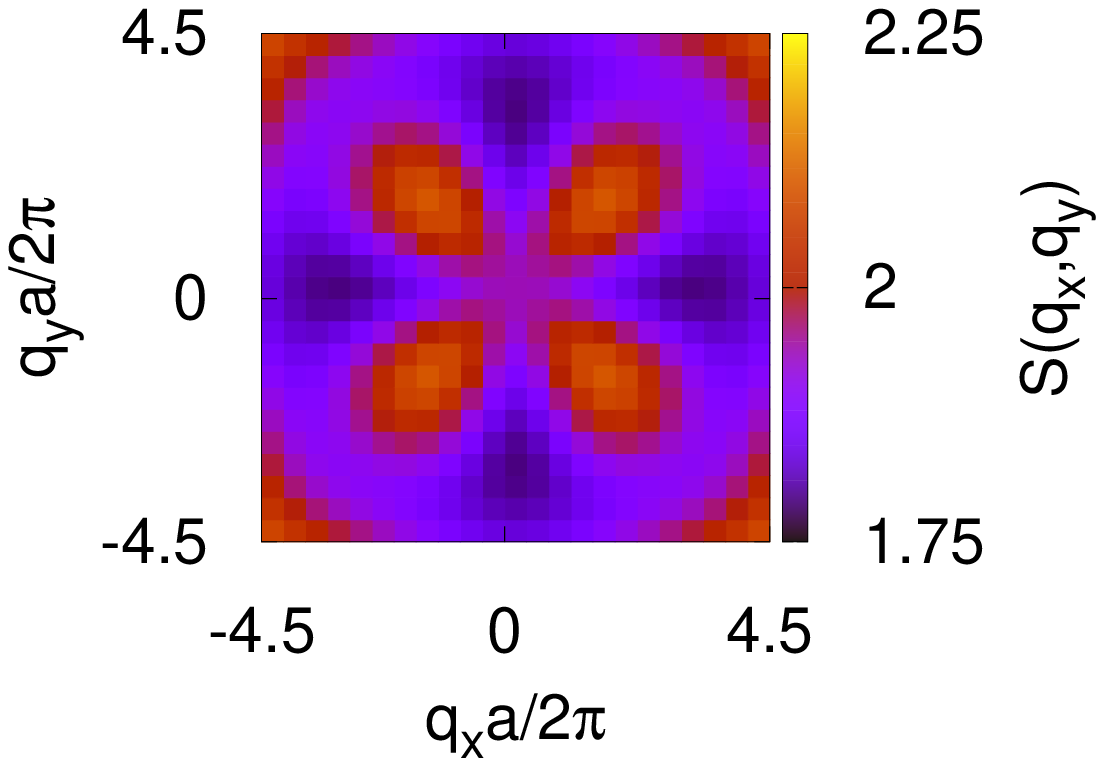}
        \caption{}
    \end{subfigure}
    \begin{subfigure}[b]{0.65\columnwidth}
        \centering
        \hspace{-0.0cm}\includegraphics[width=\columnwidth]{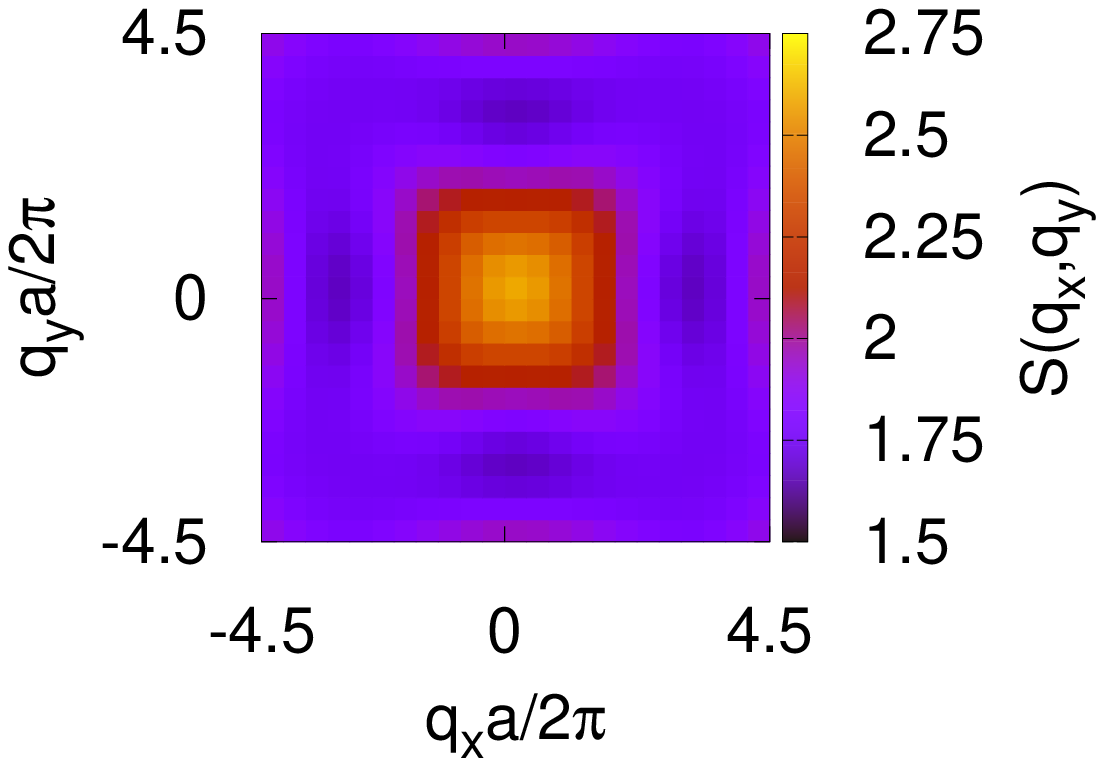}
        \caption{}
    \end{subfigure}
    \begin{subfigure}[b]{0.65\columnwidth}
        \centering
        \hspace{-0.0cm}\includegraphics[width=\columnwidth]{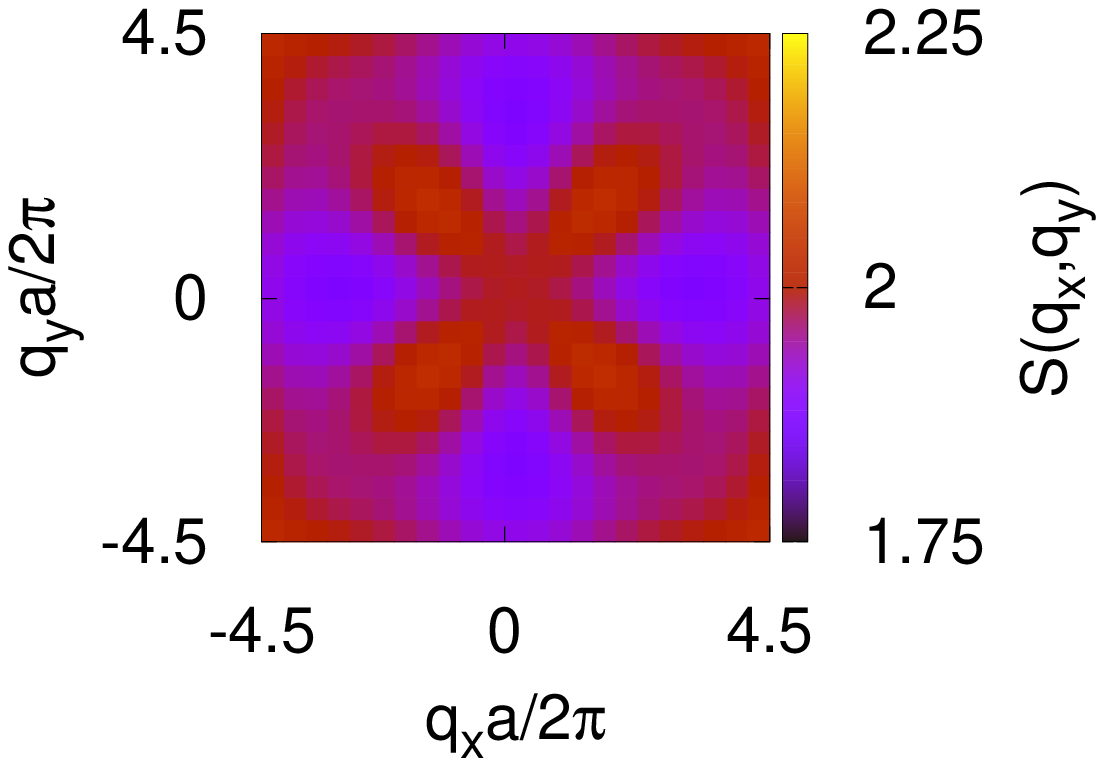}
        \caption{}
    \end{subfigure}
    \begin{subfigure}[b]{0.65\columnwidth}
        \centering
        \hspace{-0.0cm}\includegraphics[width=\columnwidth]{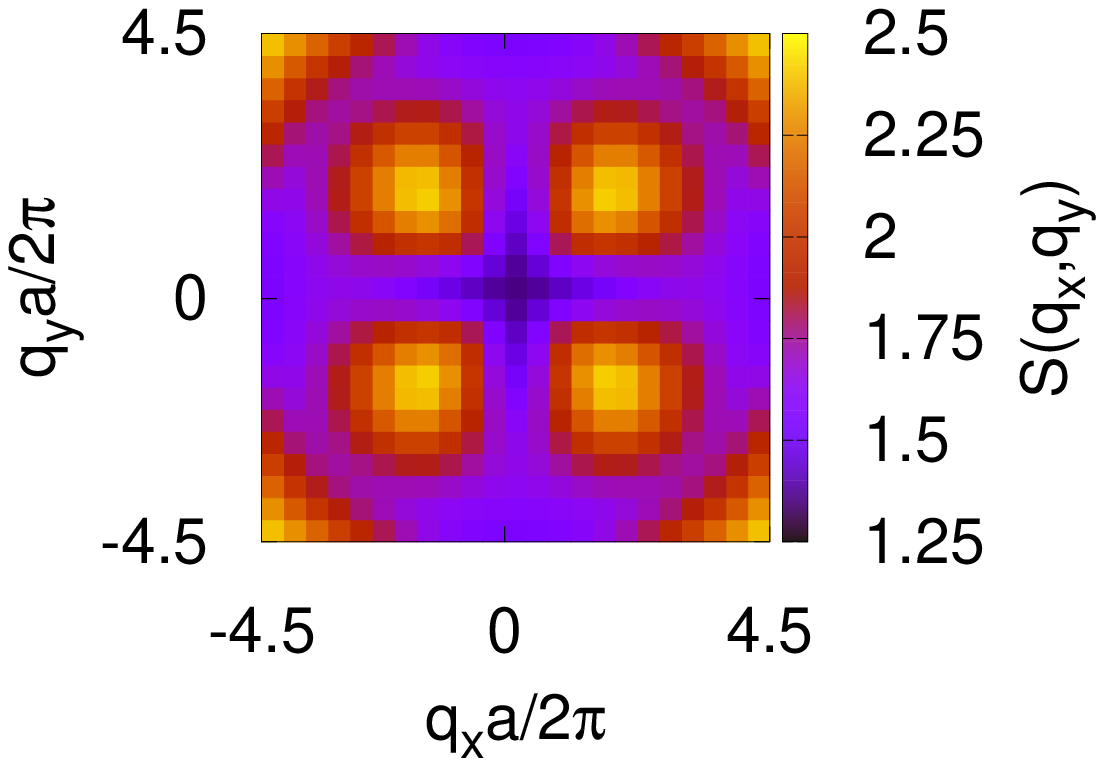}
        \caption{}
    \end{subfigure}
    \begin{subfigure}[b]{0.65\columnwidth}
        \centering
        \hspace{-0.0cm}\includegraphics[width=\columnwidth]{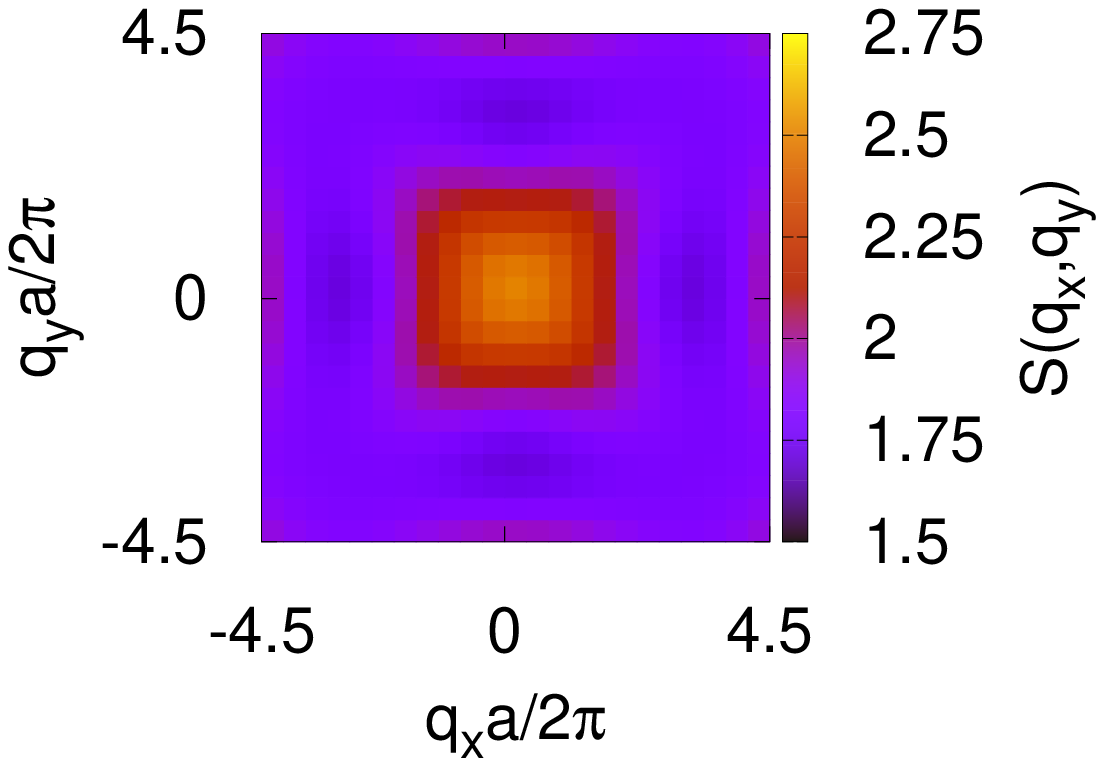}
        \caption{}
    \end{subfigure}
    \begin{subfigure}[b]{0.65\columnwidth}
        \centering
        \hspace{-0.0cm}\includegraphics[width=\columnwidth]{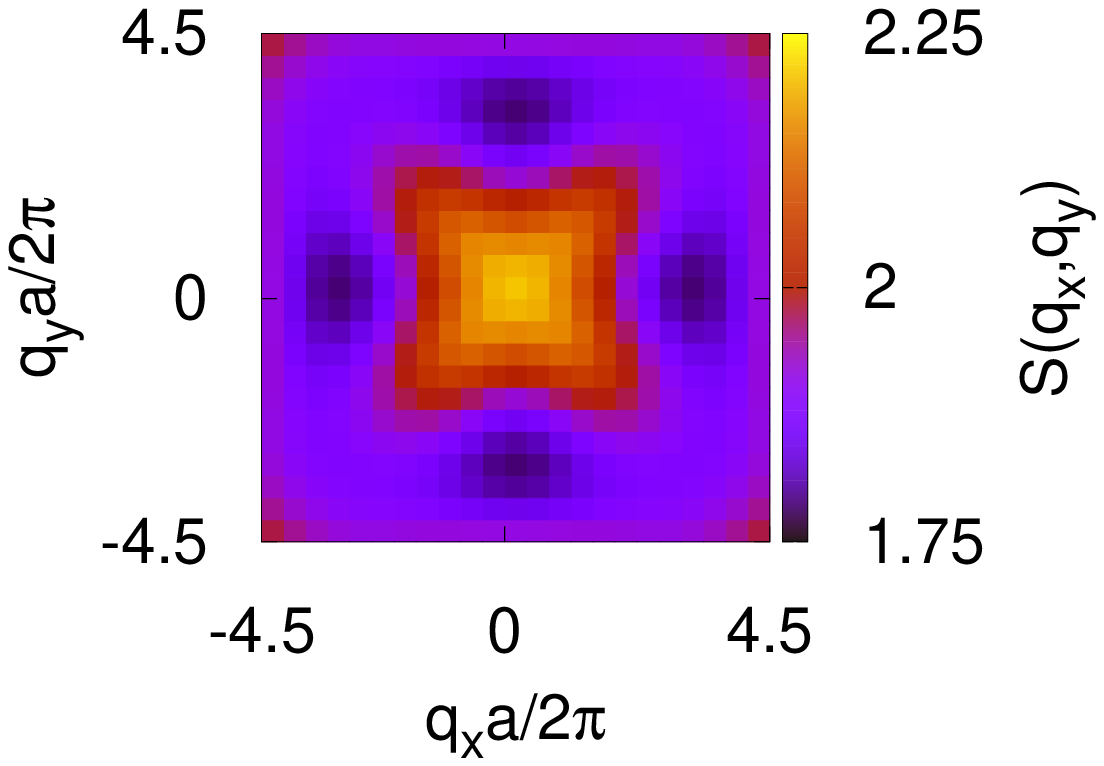}
        \caption{}
    \end{subfigure}
    \begin{subfigure}[b]{0.65\columnwidth}
        \centering
        \hspace{-0.0cm}\includegraphics[width=\columnwidth]{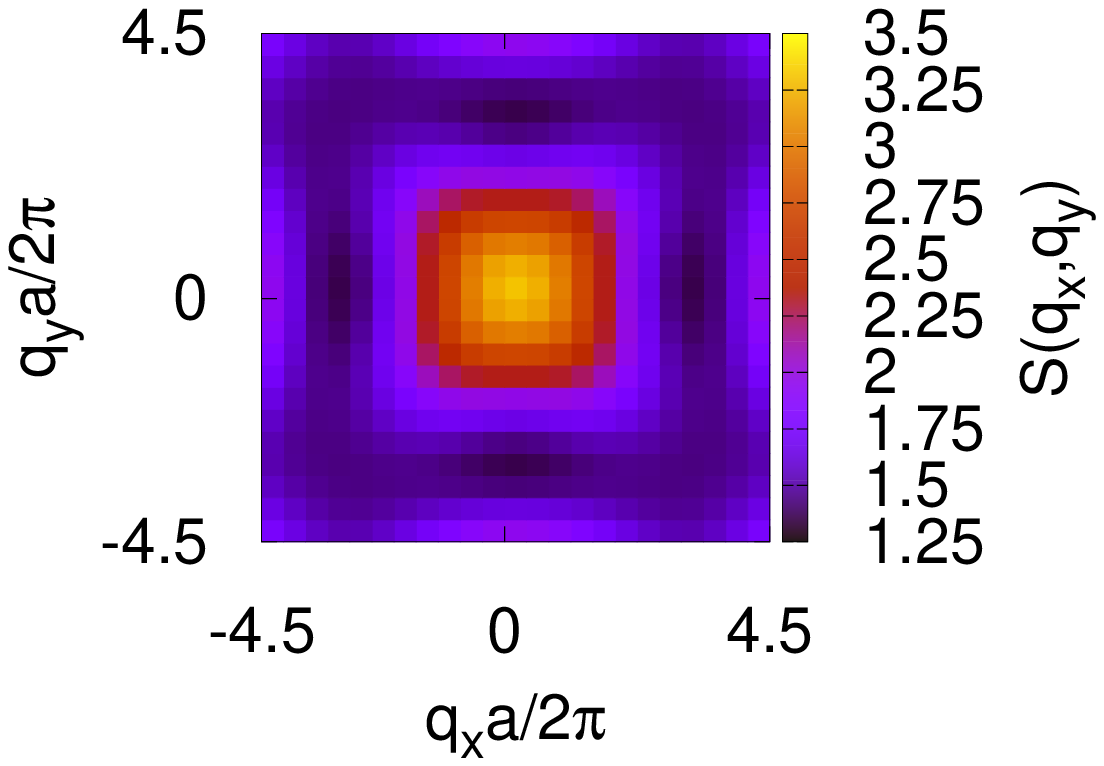}
        \caption{}
    \end{subfigure}
    \begin{subfigure}[b]{0.65\columnwidth}
        \centering
        \hspace{-0.0cm}\includegraphics[width=\columnwidth]{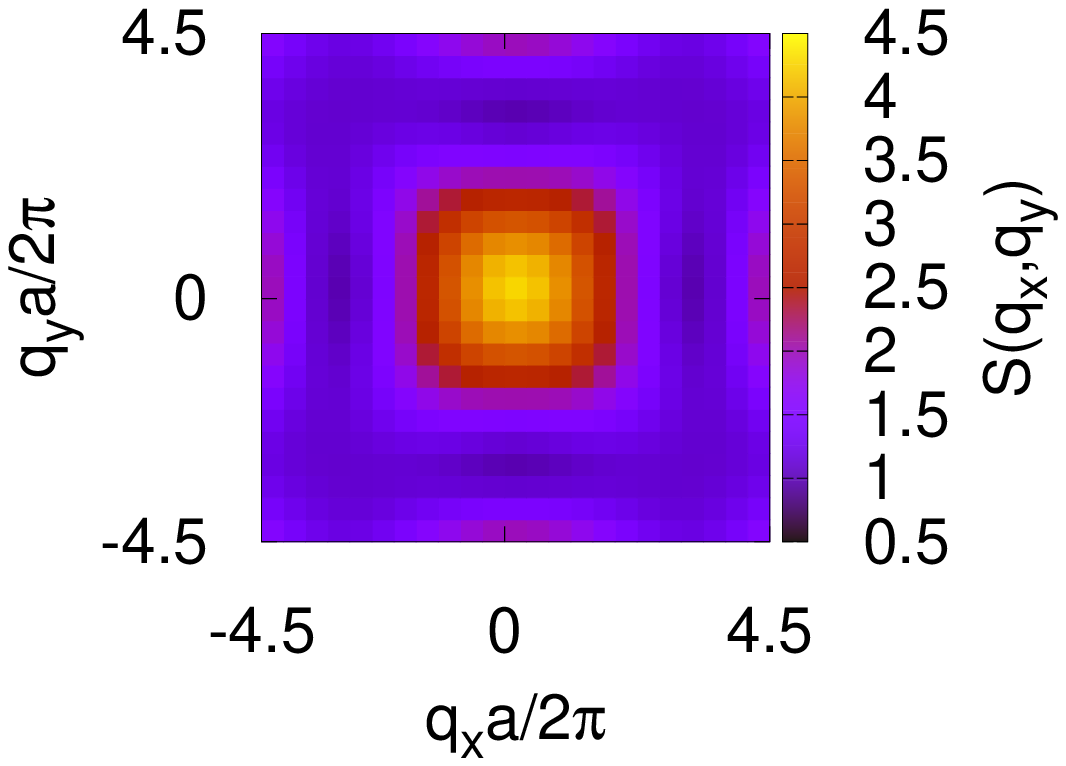}
        \caption{}
    \end{subfigure}
    \begin{subfigure}[b]{0.65\columnwidth}
        \centering
        \hspace{-0.0cm}\includegraphics[width=\columnwidth]{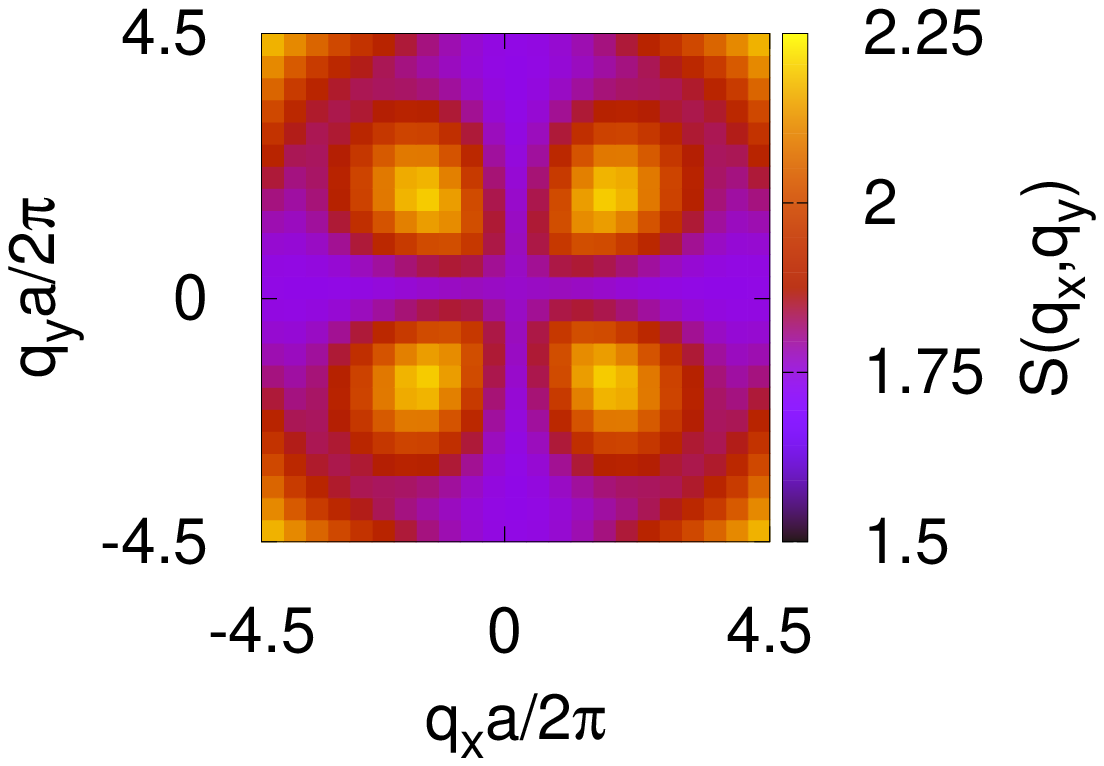}
        \caption{}
    \end{subfigure}

\caption{\label{fig:numerical_values}Training set generated using the values of magnetic field $h$ and temperature $T$ shown in Fig.~\ref{fig:size_of_training_set}~(b) and Table \ref{tab:numerical_values}. The panel labels (a-i) correspond to the index on the first column of  Table \ref{tab:numerical_values}.}
\end{figure*}

Here we give details of one of the training data sets, namely the 9-sample data set corresponding to $N=4$, $\gamma=0.6$, $\Delta=0$ and the values of temperature $T$ and magnetic field $h$ shown in Fig.~\ref{fig:size_of_training_set}~(b). Table \ref{tab:numerical_values} shows the numerical values of $h$ and $T$ for this set, along with an index a,b,c,\ldots used to identify each particular $(h,T)$ combination. Fig.~\ref{fig:numerical_values} shows the corresponding simulated neutron scattering functions. Each image constitutes one element of the training set (an ``observation''). This 9-image training set was used to obtain the principal components shown in Figs.~\ref{fig:size_of_training_set}~(e) and (h). Projecting the 9 images shown in Fig.~\ref{fig:numerical_values} onto those principal components we obtain the coordinates of the 9 filled circles shown in Fig.~\ref{fig:bifurcation_N2}~(b). The lines shown in the latter figure are obtained by projecting other images (not forming part of the training set) onto the same principal components. A similar procedure was followed when using the 3-image training set corresponding to the $(h,T)$ values in Fig.~\ref{fig:size_of_training_set}~(a) and the 500-image training set corresponding to  Fig.~\ref{fig:size_of_training_set}~(c).

\begin{table}[]
    \centering
    \begin{tabular}{c|c|c}
    Index   &    $h/J$ & $T/J$  \\\hline
a	&	0.471279	&	1.076508	\\
b	&	1.851049	&	1.753089	\\
c	&	0.528793	&	1.995304	\\
d	&	0.246644	&	0.097010	\\
e	&	1.872774	&	1.958153	\\
f	&	1.147418	&	1.732766	\\
g	&	1.750644	&	0.895347	\\
h	&	1.885494	&	0.472279	\\
i	&	0.343302	&	0.710108
    \end{tabular}
    \caption{\label{tab:numerical_values}Numerical values of the combinations of magnetic field $h$ and temperature $T$ represented graphically in Fig.~\ref{fig:size_of_training_set}~(b). The values are given in units of the exchange constant $J$, as indicated in the table headings.}
\end{table}

\section{\label{sec:individual_scores}Temperature- and field-dependence of individual principal component scores}

\begin{figure}
    \centering
    \begin{subfigure}[b]{0.85\columnwidth}
        \centering
        \hspace{-0.0cm}\includegraphics[width=\columnwidth]{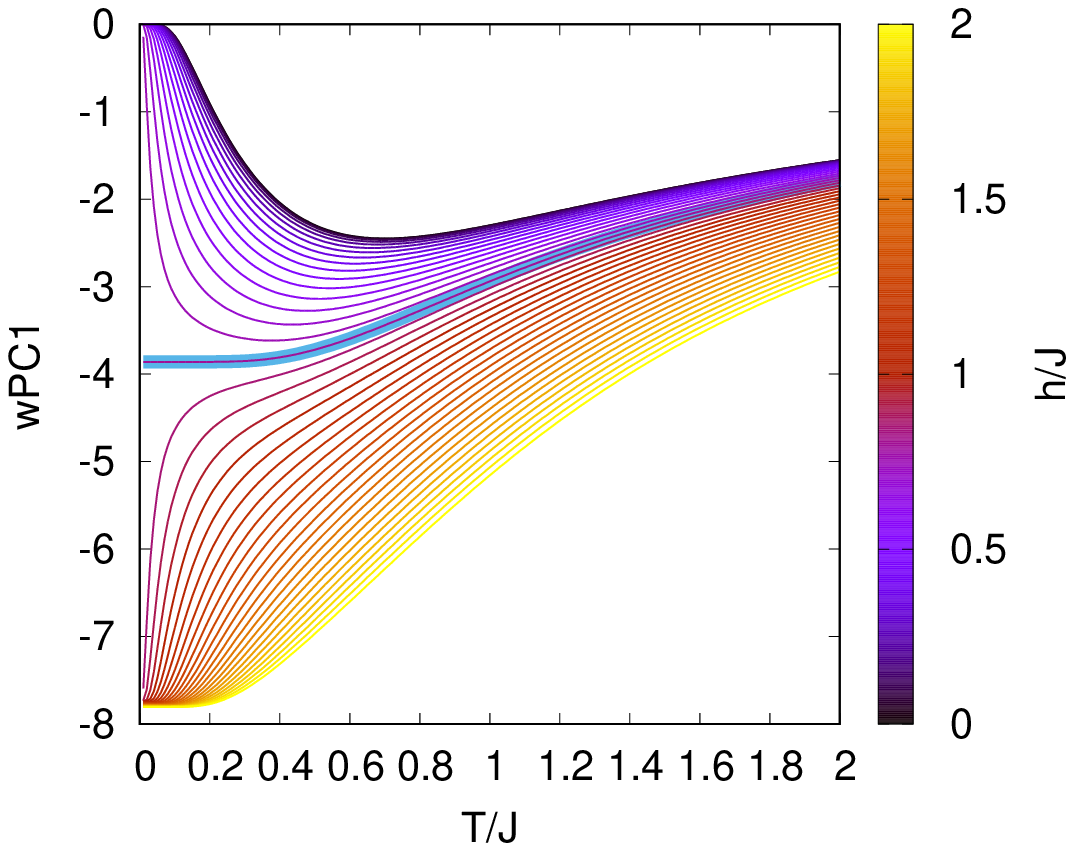}
        \caption{}
    \end{subfigure}
    \begin{subfigure}[b]{0.85\columnwidth}
        \centering
        \hspace{-0.0cm}\includegraphics[width=\columnwidth]{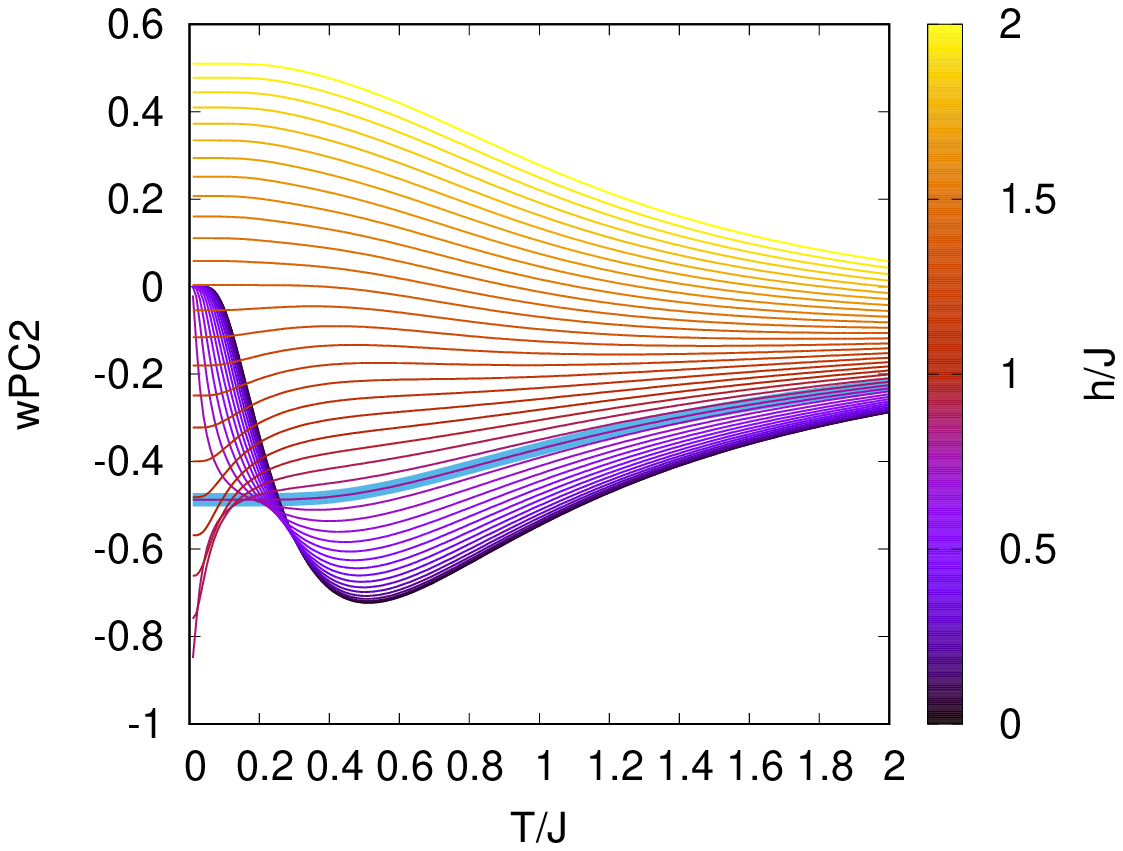}
        \caption{}
    \end{subfigure}
\caption{\label{fig:individual_scores}Temperature- and field- dependence of the individual PC scores wPC1 (a) and wPC2 (b). The data shown are from the same calculation as Fig.~\ref{fig:bifurcation_N2}~(b).}
\end{figure}

Here we discuss briefly the field- and temperature-dependences of the individual principal component scores wPC1, wPC2. These are shown in Fig.~\ref{fig:individual_scores} for the same calculation used to obtain Fig.~\ref{fig:bifurcation_N2}~(b). As we can see the temperature-dependence of the PC scores changes qualitatively at the factorisation field - for instance at low temperatures the derivative of wPC1 with respect to $T$ is positive for $h>h_f$ and negative for $h<h_f$. At exactly $h=h_f$ (highlighted in cyan) this derivative vanishes in the limit $T\to 0$. We note, however, that in a general situation we will not have a constant-field scan taken at exactly $h=h_f$ (see footnote \ref{foot:gamma_choice_footnote} on page \pageref{foot:gamma_choice_footnote}). Because of this, in general locating the value of the field where the PC scores become temperature-independent involves an extrapolation. This must be contrasted with the score bifurcation plot where wPC1 and wPC2 are shown in combination and the factorisation field can be determined precisely from finite-temperature data.

\section{\label{sec:geometric_construction}Geometric construction for determining the factorisation field from finite-temperature data}

\begin{figure*}
\includegraphics[width=\columnwidth]{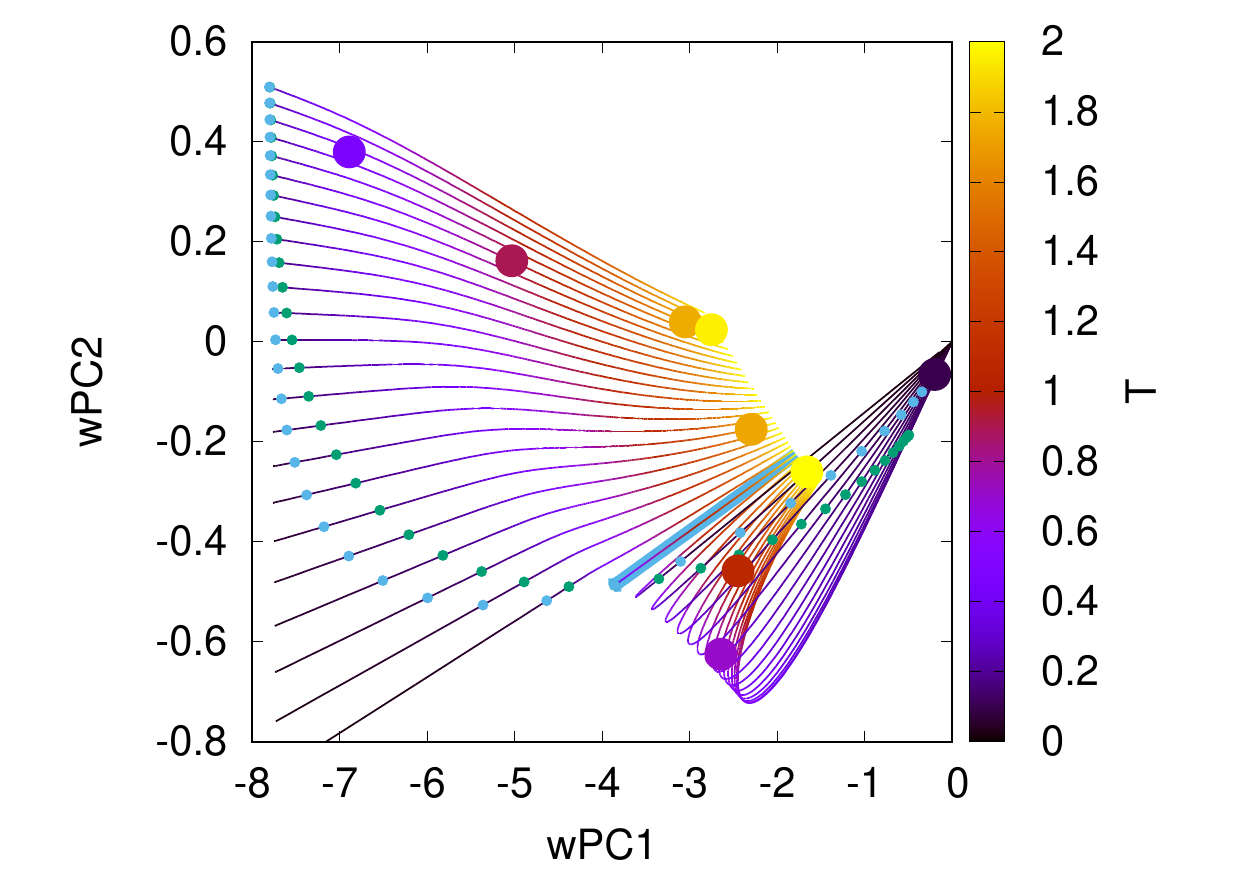}
\caption{\label{fig:bifurcation_N2_construction}Here we reproduce the plot in Fig.~\ref{fig:bifurcation_N2} (b) but with the addition of green and blue circles marking the points along the constant-field lines corresponding to temperatures $T=0.10$ and $0.15$, respectively.}
\end{figure*}

Here we describe a simple geometric construction for determining precisely the location of the factorisation field from finite-temperature data. We illustrate this by focusing on the $N=2$ data in Fig.~\ref{fig:bifurcation_N2}. The same data is reproduced in Fig.~\ref{fig:bifurcation_N2_construction}, but with additional symbols showing the points along the constant-field curves where the temperature reaches two particular values, namely $T=0.10$ and $T=0.15$. By inspection of that plot it is quite clear that the two curves will cross at the point where the bifurcation takes place, whose precise coincidence with the factorisation field we have determined numerically. It is evident that this construction will work, and give the same value, irrespective of which two temperatures we choose, as long as they are low enough to show the inflection of the curve to the right of the bifurcation. A similar construction can be made for the $N=4$ plots in Figs.~\ref{fig:bifurcation_N2_construction} (c) and \ref{fig:bifurcation_N4_lower_res}.

\bibliographystyle{apsrev4-1}
\bibliography{00_bibliography_jq}

\end{document}